# On optimum left-to-right strategies for active context-free games

Henrik Björklund *
Umeå University
henrikb@cs.umu.se

Martin Schuster
TU Dortmund
martin.schuster@udo.edu

Thomas Schwentick †
TU Dortmund
thomas.schwentick@udo.edu

Joscha Kulbatzki
SCISYS Deutschland GmbH
Joscha.Kulbatzki@scisys.de

## ABSTRACT

*Active context-free games* are two-player games on strings over finite alphabets with one player trying to rewrite the input string to match a target specification. These games have been investigated in the context of exchanging Active XML (AXML) data. While it was known that the rewriting problem is undecidable in general, it is shown here that it is EXPSPACE-complete to decide for a given context-free game, whether all safely rewritable strings can be safely rewritten in a *left-to-right* manner, a problem that was previously considered by Abiteboul et al. Furthermore, it is shown that the corresponding problem for games with finite replacement languages is EXPTIME-complete.

## 1. INTRODUCTION

In this paper, we study *Active Context-Free Games*, played by two players on finite strings. The motivation for these games comes from the study of *Active XML* documents [1, 3, 7]. In such documents, only some of the data is explicitly given, the rest of the data can be obtained by calls to Web services. An example could be a document that includes as a part the latest news headlines. Rather than storing these headlines on the host web server, a web service run by a news agency is called each time the document is requested by a user. The headlines retrieved by the call are then incorporated into the document before it is sent to the user. It can also be the case that the news agency returns another active document, i.e., one that contains further possibilities for calling web services.

However, this approach raises some challenges when documents should be valid with respect to some schemas. The hosts not only need to ensure that their own documents conform to the schema, but also that this is the case for all possible documents resulting from web service calls.

This scenario was studied by Milo et al. [7], who formulated a polynomial time algorithm for a restricted setting, in which this *schema rewriting problem* on AXML trees can be solved by recursively solving a similar rewriting problem on strings.

In order to model and study this scenario, active context-free games, or context-free games for short, were introduced by Muscholl et al. [8]. Context-free games are two-player games played on strings, where the first player, JULIET, represents the host. By calling on letters she tries to rewrite the string into one that conforms to a schema, represented by a regular language. Her opponent, ROMEO, gets to pick a string from a regular set to replace the letter JULIET called on. Starting from a given string, representing the active document, JULIET wins if the string is ever rewritten into a word in the schema language. Otherwise, ROMEO wins.

In this particular paper we focus on so-called *left-to-right* (L2R) strategies. If JULIET follows such a strategy, she is not allowed to call a position that is to the left of a previously called position. L2R-strategies have been considered before, e.g. in [8, 2]. They are more feasible than unrestricted strategies. For instance, while it is in general impossible to determine, given a game and an input string, whether JULIET has a winning strategy, it can be decided in EXPTIME whether she has a winning L2R-strategy [8]. The aforementioned efficient algorithm by Milo et al. [7] also requires a restriction to L2R rewritings.

It is thus useful to determine during the design phase of a system whether for JULIET, L2R-strategies are *universal*.[1] This is the L2RALL problem, studied here: given a game, does JULIET have a winning L2R-strategy for every string for which she has a winning strategy at all? The L2RALL problem was first considered in [2], where it was claimed to be undecidable.

We take the following approach to the problem. First we show that if L2R-strategies are not universal, there is a string for which JULIET has a winning strategy with one *left*

*We acknowledge the financial support of the Swedish Research Council through its grant 621-2011-6080.

†We acknowledge the financial support of the Future and Emerging Technologies (FET) programme within the Seventh Framework Programme for Research of the European Commission, under the FET-Open grant agreement FOX, number FP7-ICT-233599.

---

[1] The high complexity of lower bounds we prove for the L2RALL problem may seem to make this task forbiddingly difficult; however, since L2RALL is effectively a static analysis problem, the added complexity may be affordable as a pre-processing step.

*step* but no L2R winning strategy. Then we show how to construct automata for all strings with a winning L2R strategy and for all strings with a winning "1-left-step" strategy, respectively. The L2RALL problem then boils down to a containment test for these two automata. To show that the automata can be effectively (and optimally efficiently) computed, we use the concept of *effects* of a string. In a nutshell, the effect of a substring summarizes how the string that is obtained from it during the game can affect the automaton for the schema.

We show that the L2RALL problem can be solved in exponential space and that this is optimal. If the set of possible replacement strings from which ROMEO can choose is finite and explicitly given, for every letter, the complexity drops to exponential time. Thus, we prove the following result.

THEOREM 1 (MAIN THEOREM).

*(a) L2RALL is EXPSPACE-complete.*

*(b) If all replacement languages are finite and explicitly given in the input, L2RALL is EXPTIME-complete.*

The paper is organized as follows. After some preliminaries, we show in Section 3 that to decide the L2RALL problem, general strategies can be replaced by "1-left-step" strategies. In Section 4 we define effects and state their basic properties. Section 5 shows how to define and compute automata for the set of words with winning L2R strategies. In Section 6 we give the decision algorithms and in Section 7 the matching lower bounds.

**Related work.** We already discussed the most important related papers [2, 8, 1, 7] above. That automata for the set of words with winning L2R strategies can be constructed in exponential time was already shown in [8]. However, the proof did not give an explicit construction but was by reduction to algorithmic problems for pushdown systems. That L2RALL is decidable was already claimed in the Diploma thesis of Joscha Kulbatzki, which was written under the supervision of the third author [6].

We thank the anonymous reviewers of ICDT 2013 for helpful suggestions and Ahmet Kara for proof reading.

## 2. PRELIMINARIES

In this section we define the fundamental notions.

### 2.1 Context-free games

A *context-free game* $G = (\Sigma, R, T)$ consists of a finite alphabet $\Sigma$, a *rule set* $R \subseteq \Sigma \times \Sigma^*$ and a regular *target language* $T \subseteq \Sigma^*$. It is required that for each symbol $f \in \Sigma$, the set $R_f =_{\text{def}} \{u \mid (f, u) \in R\}$ is regular. By $\Gamma$ we denote the set $\Gamma =_{\text{def}} \{f \mid f \in \Sigma, R_f \neq \emptyset\}$ and we call the symbols from $\Gamma$ *function symbols*. We denote function symbols by $f, f_1, \ldots$ and *terminal symbols* from $\Sigma \setminus \Gamma$ by $a, b, a_1, \ldots$.

A play of the game $G$ is played by two players, JULIET and ROMEO, on a word $w \in \Sigma^*$.

In its original form, as introduced in [8], the game proceeds in rounds, in each of which JULIET selects a position of the current string and ROMEO chooses a rewriting rule to replace the current symbol $f$ at that position by a string from $R_f$. For the purposes of this paper a different, but equivalent, definition of (the rules of) context-free games is more suitable,

In our definition, a play can have several passes in which the focus is moved along the current string, from left to right. In each round, JULIET selects whether the current symbol in the current word should be rewritten or passed over. If she chooses a rewrite, then ROMEO chooses a substitution for the symbol that is allowed by the rule set.

More formally, a *configuration* is a tuple $C = (p, u, v) \in \{1, 2\} \times \Sigma^* \times \Sigma^*$ where $p$ is the player to move (1 for JULIET and 2 for ROMEO), $uv$ is the *current word*, and the first symbol of $v$ is *the current position*.

A *winning configuration* for JULIET is a configuration $C = (p, v, \varepsilon)$ with $v \in T$.

In each configuration $(1, u, v)$ with $v \neq \epsilon$, JULIET can either choose a Read move or, if the first symbol $f$ of $v$ is from $\Gamma$ a Call move. If she selects Read, the play moves one step to the right. If she selects Call, then ROMEO selects a string from the set $R_f$. In a configuration $(1, u, \epsilon)$ JULIET can either do a *left step* or stop the game.

A *move of* JULIET is thus represented by Read, Call, LS or Stop and a *move of* ROMEO is represented by a string $x$.

The configuration $C' = (p', u', v')$ is a *possible successor configuration* of $C = (p, u, v)$ (Notation: $C \to C'$) if

(1) $p' = p = 1$, $u' = us$, and $sv' = v$ for some $s \in \Sigma$ (JULIET plays Read);

(2) $p = 1$, $p' = 2$, $u' = u$, and $v' = v$ (JULIET plays Call);

(3) $p = 2$, $p' = 1$, $u' = u$, $v = fx$ for some $f \in \Gamma$, $v' = yx$ for some $y \in R_f$ (ROMEO plays $y$);

(4) $p' = p = 1$, $u \notin T$, $v = \varepsilon$, $v' = u$, $u' = \varepsilon$, (JULIET plays LS).

If JULIET plays Stop in a configuration $C = (p, u, \epsilon)$ we write $C \to \top$ if $u \in T$ and $C \to \bot$ if $u \notin T$ and we thus consider $\top$ and $\bot$ as configurations as well.

Since we will mostly consider configurations where JULIET is to move, we often omit the player when talking about them. Thus $(u, v)$ is a shorthand for $(1, u, v)$.

The *initial configuration* of game $G$ for string $u$ is defined as $C_0(u) =_{\text{def}} (1, \varepsilon, u)$.

A *play* of the game $G$ is either an infinite sequence $\Pi = C_0, C_1, \ldots$ or a finite sequence $\Pi = C_0, C_1, \ldots, C_k$ of configurations, where, for each $i > 0$, $C_{i-1} \to C_i$. If the sequence is finite, then $C_k$ must be either $\top$ or $\bot$. If $C_k = \top$, JULIET *wins* the play, in all other cases, ROMEO wins. We write $\Pi \equiv p$ if player $p$ wins $\Pi$.

We assume in this paper that a game $G = (\Sigma, R, T)$ is represented by a DFA $A(T)$ for $T$ and by a NFA $A_f$ for $R_f$, for every $f \in \Gamma$.[2] In the sequel, let $A(T) = (Q, \Sigma, \delta, F, q_0)$ with state set $Q$, transition function $\delta : Q \times \Sigma \to Q$, accepting states $F \subseteq Q$ and initial state $q_0 \in Q$.

We note that our definition of active context-free games is indeed equivalent to the one in [8]. JULIET can select an arbitrary position by playing a sequence of Read moves possibly followed by a LS move, another sequence of Read moves and, eventually, a Call move at the desired position.

---

[2]We note that whether $R_f$ is represented by DFAs or NFAs does not influence the complexity. However, we conjecture that allowing NFAs for $T$ may lead to an unavoidable exponential blowup of the complexity. We chose DFAs for our setting as we are interested in cases with reasonable efficiency.

## 2.2 Game trees

The *game tree* $Tree_{G,u}$ for $G$ on string $u$ is a tree labeled by configurations. Each branch of the tree represents one possible play of the game. The root of $Tree_{G,u}$ is labeled by the initial configuration $C_0(u)$. A node labeled $C$ has one child for every configuration $C'$ such that $C \to C'$. This means that the only leaves of $Tree_{G,u}$ are nodes labeled by final configurations of finite plays. In general, nodes labeled by configurations $C = (1, u, v)$ have one or two children: if $v = sv'$ for some $s \in \Sigma$, there is always one child corresponding to a Read move, and a second one corresponding to a Call move exists iff $s \in \Gamma$. If $v = \epsilon$, the two children correspond to a LS and Stop move respectively. Nodes labeled by configurations where Romeo is to move can have infinitely many children.

## 2.3 Strategies

A *strategy* for player $p \in \{1, 2\}$ maps prefixes $C_0, C_1, \ldots, C_k$ of plays, where $C_0$ is an initial configuration and $C_k$ is a $p$-configuration, to allowed moves. A strategy $\sigma$ is *memoryless* if, for every prefix $C_0, C_1, \ldots, C_k$ of a play, $\sigma(C_0, C_1, \ldots, C_k)$ only depends on $C_k$.

We denote strategies for Juliet by $\sigma, \sigma', \sigma_1, \ldots$ and strategies for Romeo by $\tau, \tau', \tau_1, \ldots$.

For configurations $C, C'$ and strategies $\sigma, \tau$ we write $C \xrightarrow{\sigma,\tau} C'$ if $C'$ is the unique successor configuration of $C$ determined by the strategies $\sigma$ and $\tau$. Given an initial word $u$ and strategies $\sigma, \tau$ the play[3] $\Pi(\sigma, \tau, u) =_{def} C_0(u) \xrightarrow{\sigma,\tau} C_1 \xrightarrow{\sigma,\tau} C_2 \cdots$ is uniquely determined.

A strategy $\sigma$ for Juliet is *finite* on string $u$ if the play $\Pi(\sigma, \tau, u)$ is finite for every strategy $\tau$ of Romeo. It is a *winning strategy* for $u$ if $\Pi(\sigma, \tau, u) \equiv 1$, for every $\tau$. A strategy $\tau$ for Romeo is a *winning strategy* for $u$ if $\Pi(\sigma, \tau, u) \equiv 2$, for every strategy $\sigma$ of Juliet.

We are particularly interested in restricted kinds of strategies of Juliet.

A *left-to-right (L2R)* strategy for Juliet is a strategy in which Juliet never does a LS move.

We denote the set of all unrestricted strategies for Juliet in the context-free game $G$ by $STRAT(G)$, and the set of all L2R-strategies by $STRAT_{L2R}(G)$. The set of all strategies for Romeo is denoted by $STRAT_{Romeo}(G)$.

By definition, $STRAT_{L2R}(G) \subseteq STRAT(G)$.

By $safe(G)$ we denote the set of all words for which Juliet has a winning strategy and by $safe_{L2R}(G)$ the set of all words for which she has a winning L2R-strategy.

In this paper we are mainly interested in the following algorithmic problem: given a context-free game $G$, decide whether $safe_{L2R}(G) = safe(G)$. By L2RALL we denote the set of all games $G$, for which $safe_{L2R}(G) = safe(G)$.

As context-free games are reachability games we can make use of the following classical result; see, e.g., [5].

THEOREM 2. *Let $G$ be context-free game, and $u$ a string. Then the following statements holds for the game starting from $u$.*

(a) *Either Juliet or Romeo has a winning strategy. If Juliet or Romeo has a winning strategy then they also have a memoryless strategy.*

---
[3]As the underlying game $G$ will always be clear from the context, our notation does not mention $G$ explicitly.

(b) *Either Juliet has a winning L2R strategy or Romeo has a winning strategy against all L2R strategies. If Juliet has a winning L2R strategy then she also has a memoryless winning L2R strategy. If Romeo has a winning strategy against all L2R strategies then he also has a memoryless such strategy.*

Therefore, we will only consider memoryless strategies. Thus, in the following, strategies $\sigma$ for Juliet map configurations $C$ to moves $\sigma(C) \in \{Call, Read\}$ and strategies $\tau$ for Romeo map configurations $C$ to moves $\tau(C) \in \Sigma^*$.

We sometimes consider *subgames* on a certain part of a string and talk about strategies for subgames. From a configuration $(u, vw)$, Juliet can use a strategy $\sigma$ on the subgame on $v$. This means that she follows $\sigma$ until a configuration $(uv', w)$ is reached.

The *strategy tree* for a strategy $\sigma$ of Juliet is the restriction $Tree_{G,u}(\sigma)$ of $Tree_{G,u}$ to $\sigma$. In other words, for nodes labeled by configurations where Juliet is to move, we remove all subtrees rooted at children labeled by configurations that are not selected by $\sigma$. Strategy trees for Romeo are defined symmetrically. If we fix strategy $\sigma$ for Juliet and $\tau$ for Romeo, we get $Tree_{G,u}(\sigma, \tau)$, which only has one branch, labeled by the play $\Pi(\sigma, \tau, u)$. Notice that if a strategy $\sigma$ of Juliet is winning, then $Tree_{G,u}(\sigma)$ has no infinite branches.

If $\Pi(\sigma, \tau, w)$ is finite, then $word^G(w, \sigma, \tau)$ is the word in the final configuration of the play on $w$ following $\sigma$ and $\tau$. (and otherwise $word^G(w, \sigma, \tau) = \bot$). We let

$$words^G(w, \sigma) =_{def} \{word^G(w, \sigma, \tau) | \tau \in STRAT_{Romeo}(G)\}.$$

As usual, if the game $G$ is clear from the context, we shall omit $G$ from the notation. We may also restrict these definitions in a natural way to only include finite or L2R-strategies where mentioned.

To deal with "game effects" the following will be useful. We call a set of sets *normal* if it does not contain two sets $X$ and $Y$ with $X \subset Y$. A finite set $S$ of finite sets can be *normalized* by applying the Norm operator, defined as follows.

$$\text{Norm}(S) = \{Y \in S \mid \text{there is no } X \in S, \text{such that } X \subset Y\}.$$

LEMMA 3. *Let $S_1, S_2$ be normal sets of sets. If for every $s_1 \in S_1$ there is $s_2 \in S_2$ such that $s_2 \subseteq s_1$ and vice versa then $S_1 = S_2$.*

PROOF. We show that every set $s_1 \in S_1$ is also in $S_2$. The lemma then follows by symmetry.

Let thus $s_1 \in S_1$. By our assumption there is $s_2 \in S_2$ such that $s_2 \subseteq s_1$ and there is a set $s'_1 \in S_1$ such that $s'_1 \subseteq s_2$. However, as $S_1$ is normal, $s_1 = s'_1$ and we get $s_1 = s'_1 \subseteq s_2 \subseteq s_1$ and thus $s_1 = s_2$. □

Where notation is dense, we sometimes just use $N(S)$ for $\text{Norm}(S)$.

## 3. FROM GENERAL TO L2R$^+$-STRATEGIES

*Definition 1.* A strategy $\sigma$ of Juliet is an *extended L2R-strategy* (L2R$^+$) if for every string $u$ and every strategy $\tau$ of Romeo, Juliet plays LS at most once and plays at most one Call before the LS-move.

LEMMA 4. *Let $G$ be a context-free game. Then $safe(G) = safe_{L2R}(G)$ if and only if $safe_{L2R^+}(G) = safe_{L2R}(G)$.*

PROOF. If $safe(G) = safe_{L2R}(G)$, then $safe_{L2R^+}(G) = safe_{L2R}(G)$ by definition.

Assume that $safe(G) \neq safe_{L2R}(G)$ and let $w$ be a string in $safe(G) \setminus safe_{L2R}(G)$. Let $\sigma$ be a winning strategy for JULIET on $w$, i.e., starting from the configuration $(1, w, \varepsilon)$. Consider the strategy tree $Tree_{G,w}(\sigma)$. In addition to the configuration labels, we mark each node $n$ in this tree with a value $LS(n)$, where $LS(n)$ is the maximum number of LS moves, on any branch of the subtree rooted in $n$. Since the tree has infinite branching, the value $LS(n)$ can, in general, be unbounded, i.e., $LS(n) = \infty$. Since $\sigma$ is a winning strategy, however, the tree has no infinite branches.

Nodes $n$ with $LS(n) \neq \infty$ and $LS(n) > 0$ are also marked by $Calls(n)$, the maximum number of Call moves that occur before the first LS step, on any branch of the subtree rooted in $n$. We note that $Calls(n)$ might be $\infty$.

In the following, we call, for nodes $n$ with $LS(n) \neq \infty$, the pair $(LS(n), Calls(n))$ the *marking* of $n$ and we denote by $\leq$ the lexicographic order on markings.

Without loss of generality, we may assume that $\sigma$ is *optimally efficient* in the following sense. We assume that for every node $n$ of the strategy tree, labeled with a configuration $(p, u, v)$, such that $LS(n) \neq \infty$, there is no other winning strategy $\sigma'$ on $w$, such that the strategy tree for $\sigma'$ and $w$ has a node $n'$ labeled with the same configuration but having a lexicographically smaller marking. Such an optimally efficient strategy can be constructed for every configuration $(p, u, v)$ by nested induction on the minimal value of $(LS(n), Calls(n))$ that nodes $n$ representing $(p, u, v)$ can assume in winning strategies for $(p, u, v)$.

As $safe(G) \neq safe_{L2R}(G)$, there must be a node $n$ in $Tree_{G,w}(\sigma)$ with $LS(n) > 0$.

We first show that $Tree_{G,w}(\sigma)$ must contain nodes $n$ with $LS(n) > 0$, $LS(n) \neq \infty$ and with a marking different from $(1,0)$, i.e. configurations in which JULIET actually has to make at least one more Call before her last LS move.

If $Tree_{G,w}(\sigma)$ has nodes with LS-value $\infty$, it also has a node $n'$, where $LS(n') = \infty$, but $LS(n) \neq \infty$, for every child node $n$ of $n'$. Otherwise, $Tree_{G,w}(\sigma)$ would have infinite branches, contradicting the fact that $\sigma$ is a winning strategy. There must be arbitrarily large LS-values among the children of $n'$ as otherwise $LS(n') \neq \infty$. In particular, $n'$ must have a JULIET-grandchild $n$ with $LS(n) > 1$ and therefore a marking differing from $(1,0)$.

If $Tree_{G,w}(\sigma)$ has no nodes with LS-value $\infty$, then for the root $r$ of $Tree_{G,w}(\sigma)$ it holds $LS(r) \neq \infty$, and thus $LS(r) \geq 1$ (as otherwise $w \in safe_{L2R}(G)$) and $Calls(r) > 0$ (as otherwise one LS-step less would suffice — at the root the current position is 1!).

Thus, there must be a JULIET-node $n_1$ with $LS(n_1) > 1$, $LS(n_1) \neq \infty$ and with a marking different from $(1,0)$.

Let $n$ be any node with $LS(n) > 0$, $LS(n) \neq \infty$ and with a marking $(i, j) \neq (1, 0)$. For the markings of the children and grandchildren of $n$ there are the following possibilities.

(i) JULIET plays Read on $n$ and for the unique child $n'$ of $n$ the marking is $(i, j)$.

(ii) JULIET plays Call on $n$, $j = \infty$, and there is a grandchild $n'$ of $n$ with marking $(i, \infty)$.

(iii) JULIET plays Call on $n$, $j = \infty$, there are grandchildren $n''$ with $LS(n'') = i$ and for all grandchildren markings of the form $(i, j')$, $j' \neq \infty$. In particular, there is a grandchild $n'$ with marking $(i, j')$, for some $j' > 0$.

(iv) JULIET plays Call on $n$, $j \neq \infty$, and all grandchildren have markings that are strictly smaller than $(i, j)$, including one child $n'$ with marking $(i, j - 1)$.

(v) JULIET plays LS on $n$, $j = 0$ and the child $n'$ of $n$ has a configuration of the form $(1, u, \varepsilon)$ and marking $(i - 1, j')$ with $j' > 0$.

We can construct a sequence $n_1, n_2, \ldots$ of nodes by chosing, in all cases (i)-(v), $n_{i+1} = n'_i$, for $i \geq 1$. As this sequence follows a branch of the tree and $n_1$ is a winning node for $\sigma$, the sequence can not be infinite. Furthermore, each leaf has marking $(1, 0)$. Therefore, the sequence must contain a JULIET-node $n_\ell$ with marking $(1, 1)$. Let $(1, x, y)$ be the configuration of $n_\ell$. We claim that $xy \in safe_{L2R^+}(G) \setminus safe_{L2R}(G)$.

First, $xy \notin safe_{L2R}(G)$, as otherwise the marking of $n_\ell$ would be at most $(1, 0)$ (no Call move needed before the LS-step).

On the other hand, as the marking of $n_\ell$ is $(1, 1)$, starting from $(1, xy, \varepsilon)$, JULIET can play Read on $x$ and can win with one Call before the one and only LS move, therefore $xy \in safe_{L2R^+}(G)$.

Thus, $safe_{L2R^+}(G) \neq safe_{L2R}(G)$, completing the proof. □

## 4. EFFECTS FOR L2R STRATEGIES

*Effects* are a way to summarize the impact with respect to the automaton $A(T)$ of the possible strings by which a (sub-)string can be rewritten in one pass of a play. (Recall that $A(T) = (Q, \Sigma, \delta, F, q_0)$ is the DFA accepting the regular language $T$.) In this section, we only consider L2R strategies for JULIET, that is, JULIET never makes an LS-move.

Suppose we have the game configuration $(1, v, uw)$. As play goes on, it will eventually reach a configuration $(1, vu', w)$, where $u$ has been traversed and rewritten into $u'$. If we fix a strategy for JULIET and ROMEO then $u'$ is uniquely determined (unless the subgame on $u$ does not terminate). If we only fix a strategy $\sigma$ for JULIET, each strategy of ROMEO determines a string $u'$ (or does not terminate) and we can associate the set $words(u, \sigma)$ with $\sigma$. The *relative effect* $e(\sigma, u, q)$ of $u$ for a strategy $\sigma$ of JULIET and a state $q$ is just the set of states that $A(T)$ can reach by reading strings in $words(u, \sigma)$, starting from state $q$. The *effect* of $u$ is basically the set of all such sets $e(\sigma, u, q)$, for all states $q$ and strategies $\sigma$.

Thus $E[u]$ is a mapping that assigns to every state of $Q$ a *set of sets* of states and thus its type is $Q \to \mathcal{P}(\mathcal{P}(Q))$.

*Definition 2.* Let $u$ be a string, $q \in Q$ a state and $\sigma$ a L2R-strategy of JULIET. The *relative* effect $e(\sigma, u, q)$ is the set $\{\delta^*(q, w) | w \in words(u, \sigma)\}$ or $\perp$ if $\perp \in words^G(w, \sigma)$.

The effect $E[u]$ of $u$ maps every state $q$ to the normalized set of relative effects $e(\sigma, u, q)$ of $u$ for all $\sigma \in \text{STRAT}_{L2R}$.

Stated less formally, $e(\sigma, u, q)$ is the set of states for which there is a strategy $\tau$ of ROMEO and a string $w \in \Sigma^*$ such that $w = word(u, \sigma, \tau)$ and $\delta^*(q, w) = p$, or $\perp$ if $\perp \in words^G(w, \sigma)$. The definition of the effect $E[u]$ uses normalized sets of relative effects as JULIET can always restrict herself to strategies with minimal relative effects.

LEMMA 5. *Let $u$ be a string and $G$ a context-free game. Then, $u \in safe_{L2R}(G)$ if and only if there is a relative effect $e \in E[u](q_0)$ for which $e \subseteq F$.*

PROOF. The latter condition is equivalent to the existence of a strategy for JULIET for which all states that can be reached by counter-strategies of ROMEO are in $F$ and therefore is equivalent to $u \in \text{safe}_{L2R}(G)$. □

If we want to stress the game relative to which an effect is defined, we add a superscript to this notation as in $E^G[s]$ or in $e^G(\sigma, s, q)$.

It should be noted that strategies of JULIET for which ROMEO has a non-terminating counter strategy are not reflected in the effect of a word $u$. We tacitly assume that JULIET will always follow a strategy that guarantees termination (and such strategies are always available as JULIET can simply stick to Read moves).

Henceforth, we will often consider relative effects and effects without having an underlying word $u$ at hand. An (abstract) relative effect is just an element of $\mathcal{P}(Q)$. An (abstract) effect is a mapping $E$ of type $Q \to \mathcal{P}(\mathcal{P}(Q))$, such that every $E[q]$ is normal. We denote the set of all[4] abstract effects by $\mathcal{E}$.

*Composition.*

We next define the composition operation $\circ$ for effects. If $E_1 = E[u]$ and $E_2 = E[v]$ then $E_1 \circ E_2$ should just be $E[uv]$. However, we need a definition of $\circ$ for abstract effects, that is, a definition that is independent of the strings $u$ and $v$.

The definition uses the operation MIX, which is defined on sets of sets of sets. Let $\mathcal{D} = \{D_1, \ldots, D_n\}$ be a set of sets of sets. Then $\text{MIX}(\mathcal{D})$ is the set

$$\text{NORM}(\{d_1 \cup \cdots \cup d_n \mid d_1 \in D_1 \wedge \cdots \wedge d_n \in D_n\}).$$

In other words, the MIX operation computes every way of taking the union of one element from each of $D_1, \ldots, D_n$.

We define the composition $E_1 \circ E_2$ of two abstract effects $E_1, E_2 : Q \to \mathcal{P}(\mathcal{P}(Q))$ as follows.

$$(E_1 \circ E_2)(q) = \text{NORM}(\bigcup_{X \in E_1(q)} \text{MIX}(\{E_2(p) \mid p \in X\})).$$

Intuitively, for all sets $X$ that JULIET can choose from $E_1(q)$, JULIET can answer each choice of a state $p \in X$ by ROMEO with a strategy from $E_2(p)$. The resulting state sets, for each $X$ have to be put together into one set of states that JULIET can enforce by some strategy.

LEMMA 6. *Let $u, v$ be strings. Then $E[u] \circ E[v] = E[uv]$.*

PROOF. We show that, for each $q$, it holds that, for each relative effect $e$ in $(E[u] \circ E[v])(q)$ there is a relative effect $e' \in E[uv](q)$ with $e' \subseteq e$ and vice versa. The statement of the lemma then follows by minimality of relative effects.

Let $e \in (E[u] \circ E[v])(q)$ be a relative effect. We show that there is a relative effect $e' \in E[uv](q)$ such that $e' \subseteq e$.

By definition of $\circ$ there is a relative effect $X = \{q_1, \ldots, q_k\} \in E[u](q)$ and relative effects $e_2^i \in E[v](q_i)$, for each $i$, such that $e = \bigcup_{i=1}^{k} e_2^i$.

We denote the strategy of JULIET on $u$ yielding $X$ by $\sigma_1$ and the strategies on $v$ yielding $e_2^1, \ldots, e_2^k$ (from $q_1, \ldots, q_k$, respectively) by $\sigma_2^1, \ldots, \sigma_2^k$, respectively.

We define a strategy $\sigma$ on $uv$ for JULIET as follows. In the first phase, on $u$, JULIET plays according to $\sigma_1$. If $y$ is the word by which $u$ is rewritten in the game on $u$, then $\delta^*(q, y) = q_i$, for some $i \in \{1, \ldots, k\}$. In the second phase, on $v$, JULIET plays according to strategy $\sigma_2^i$.

---
[4]As always, we assume that the target automaton $A(T)$ is fixed.

We claim that for $e' = e(\sigma, uv, q)$ it holds $e' \subseteq e$. Let $p \in e'$ be arbitrarily chosen. Thus, there is a strategy $\tau$ of ROMEO such that the word $w = \text{word}(uv, \sigma, \tau)$ fulfills $\delta^*(q, w) = p$. We can write $w$ as $w_1 w_2$, where $w_1$ is the rewriting of $u$ and $w_2$ the rewriting of $v$ in the game following $\sigma$ and $\tau$.

By definition of $X = e(\sigma_1, u, q)$ and the definition of $\sigma$ it follows that $\delta^*(q, w_1) \in X$ and thus $\delta^*(q, w_1) = q_i$, for some $i$. Therefore, JULIET plays according to $\sigma_2^i$ in the game on $v$ and consequently $\delta^*(q_i, w_2) \in e(\sigma_2^i, v, q_i) = e_2^i$. Altogether,

$$p = \delta^*(q, w) = \delta^*(\delta^*(q, w_1), w_2) = \delta^*(q_i, w_2) \in e_2^i \subseteq e,$$

as required.

Next we show that for each relative effect $e' \in E[uv](q)$ there is a relative effect $e$ in $(E[u] \circ E[v])(q)$ with $e \subseteq e'$.

Let $e' \in E[uv](q)$ be a relative effect and let $\sigma$ be a strategy of JULIET such that $e' = e(\sigma, uv, q)$. Let $\sigma'$ denote the strategy of JULIET on $u$ that is induced by $\sigma$ and let $X = e(\sigma', u, q) = \{q_1, \ldots, q_k\}$.

Let $w_1, \ldots, w_k$ be words from $\text{words}(u, \sigma')$ such that, for every $i$, $\delta^*(q, w_i) = q_i$ and let $\tau_1, \ldots, \tau_k$ be corresponding strategies of ROMEO on $u$. For every $i$, let $\sigma_i$ denote the strategy of JULIET on $v$ induced by $\sigma$ from configuration $(w_i, v)$ on and let $X_i = e(\sigma_i, v, q_i)$. Finally, let

$$e = \bigcup_{i=1}^{k} X_i \in \text{MIX}(\{E[v](p) \mid p \in X\}).$$

We claim that $e \subseteq e'$: Let $p$ be an arbitrary state in $e$, thus $p \in X_i$, for some $i$. There exists a strategy $\tau'$ of ROMEO on $v$ such that for the word $z = \text{word}(v, \sigma_i, \tau')$ it holds $\delta * (q_i, z) = p$. Combining $\tau_i$ (on $u$) and $\tau'$ (on $v$) yields a strategy $\tau$ for ROMEO such that $\text{word}(uv, \sigma, \tau) = w_i z$. Furthermore, $\delta^*(q, w_i z) = \delta^*(\delta^*(q, w_i), z) = p$ and thus $p \in e' = e(\sigma, uv, q)$. □

## 5. AUTOMATA FOR L2R STRATEGIES

In this section, we define, for each context-free game $G$, NFAs $A_{\text{L2R}}(G)$ and $\hat{A}_{\text{L2R}}(G)$ for $\text{safe}_{L2R}(G)$ and $\Sigma^* \setminus \text{safe}_{L2R}(G)$, respectively. One of them, $A_{\text{L2R}}(G)$, is based on the computation of relative effects of the form $e(\sigma, u, q_0)$ for strategies $\sigma$ of JULIET, whereas the other is based on the computation of dual effects (to be defined below) of the form $\hat{e}(\tau, u, q_0)$ for strategies $\tau$ of ROMEO. Besides defining these automata and proving their correctness we also show how they can be computed in exponential time from $G$.

### 5.1 Definition and Correctness of L2R automata

*Definition 3.* Let $G = (\Sigma, R, T)$ be a context-free game with a DFA $A(T) = (Q, \Sigma, \delta, q_0, F)$ for $T$. The NFA $A_{\text{L2R}}(G) = (Q_{L2R}, \Sigma, \delta_{L2R}, \{q_0\}, F_{L2R})$ is defined as follows:

- $Q_{L2R} = \mathcal{P}(Q)$;

- $\delta_{\text{L2R}}(X, s) = \text{MIX}(\{E[s](q) \mid q \in X\})$, for each $X \subseteq Q$ and $s \in \Sigma$;

- $F_{L2R} = \mathcal{P}(F)$.

PROPOSITION 7. *Let $G = (\Sigma, R, T)$ be a context-free game with a DFA $A(T) = (Q, \Sigma, \delta, q_0, F)$ for $T$. Then $L(A_{L2R}(G)) = \text{safe}_{L2R}(G)$.*

PROOF. We show by induction on $|u|$ that for every string $u \in \Sigma^*$ we have $\text{NORM}(\delta^*_{\text{L2R}}(\{q_0\}, u)) = E[u](q_0)$:

For $u = \epsilon$, $\text{NORM}(\delta^*_{\text{L2R}}(\{q_0\}, \epsilon)) = \{\{q_0\}\} = E[\epsilon](q_0)$.

For $u = vs$ we get

$$\begin{aligned}
N(\delta^*_{\text{L2R}}(\{q_0\}, vs)) &= N(\bigcup_{X \in \delta^*_{\text{L2R}}(\{q_0\}, v)} \delta_{\text{L2R}}(X, s)) \\
&= N(\bigcup_{X \in E[v](q_0)} \delta_{\text{L2R}}(X, s)) \\
&= N(\bigcup_{X \in E[v](q_0)} \text{MIX}(\{E[s](q) \mid q \in X\})) \\
&= (E[v] \circ E[s])(q_0) \\
&= E[u](q_0).
\end{aligned}$$

We can conclude as follows that JULIET has a L2R winning strategy on $u$ if and only if $A_{\text{L2R}}(G)$ accepts $u$.

$$\begin{aligned}
u \in \text{safe}_{L2R}(G) &\Leftrightarrow \exists e \in E[u](q_0) : e \subseteq F \\
&\Leftrightarrow E[u](q_0) \cap \mathcal{P}(F) \neq \emptyset \\
&\Leftrightarrow N(\delta^*_{\text{L2R}}(\{q_0\}, u)) \cap \mathcal{P}(F) \neq \emptyset \\
&\Leftrightarrow \delta^*_{\text{L2R}}(\{q_0\}, u) \cap \mathcal{P}(F) \neq \emptyset
\end{aligned}$$

□

## 5.2 Computing L2R automata

PROPOSITION 8. *There is an algorithm that computes in exponential time the NFA $A_{L2R}(G)$ for each context-free game $G = (\Sigma, R, T)$, provided that $T$ is represented by a DFA $A(T) = (Q, \Sigma, \delta, q_0, F)$ and the sets $R_f$ are represented by DFAs, NFAs or regular expressions.*

PROOF. The algorithm first computes in exponential time the effects $E[s]$, for every symbol $s \in \Sigma$. To this end, it uses Algorithm 1 below. By Proposition 9 below, this is possible in exponential time. The construction of $A_{\text{L2R}}(G)$ is then straightforward. It should be noted that $\text{MIX}(\{E[s](q) \mid q \in X\})$ can be computed in exponential time as $|\{E[s](q) \mid q \in X\}| \leq |Q|$ and each set $E[s](q)$ is of at most exponential size. □

We next show how to compute the effect $E[s]$ for each symbol $s$ of a context-free game $G$ by a monotone fixed-point computation in exponential time. The pseudo-code of our algorithm is stated as Algorithm 1 below.

The algorithm uses a variable $P(s, q)$ for each symbol $s$ and every state $q \in Q$, intended to represent $E[s](q)$ and maintains the invariant $P(s, q) \subseteq E[s](q)$. In other words, for each set $X$ in $P(s, q)$, there is an L2R-strategy $\sigma$ of JULIET such that $X = e(\sigma, s, q)$.

Slightly abusing notation, we write $P[s]$ for the function defined by $q \mapsto P(s, q)$. It should be noted that during the computation the functions $P[s]$ need not be "real effects" in the sense that there is some string $u$ with $P[s] = E[u]$. They are rather "partial effects", that is, arbitrary functions of type $Q \to \mathcal{P}(\mathcal{P}(Q))$.

In the description of the algorithm, we use $P[w]$ as a shorthand for $P[a_1] \circ \cdots \circ P[a_\ell]$, where $a_1 \cdots a_\ell = w$ and the operation $\circ$ is defined just as for effects.

**Algorithm 1** Compute the effects of symbols from $\Sigma$

    **for all** $s \in \Sigma, q \in Q$ **do**
2:      $P[s](q) \leftarrow \{\{\delta_s(q)\}\}$
    **while** some set $P[s](q)$ has changed in the previous iteration **do**
4:      **for all** $f \in \Gamma, q \in Q$ **do**
        $P[f](q) \leftarrow P[f](q) \cup \text{MIX}(\{P[w](q) \mid w \in R_f\})$
6:      $P[f](q) \leftarrow \text{NORM}(P[f](q))$

PROPOSITION 9. *Algorithm 1 computes, for every context-free game $G = (\Sigma, R, T)$, the effect $E[s]$, for every symbol $s \in \Sigma$. Provided that $T$ is represented by a DFA $A(T) = (Q, \Sigma, \delta, q_0, F)$ and the sets $R_f$ are represented by DFAs, NFAs or regular expressions it can be carried out in exponential time.*

PROOF. We first show how the algorithm can be implemented such that it runs in exponential time. We assume without loss of generality that all sets $R_f$ are represented by NFAs.

As every set $P[s](q)$ can only contain sets of at most exponential size (in $|Q|$), the number of iterations of the while loop is at most exponential. The implementation of line 6 will make sure that $P[f](q)$ is always normal. It thus only remains to show how to implement a single execution of line 5,

$$P[f](q) \leftarrow P[f](q) \cup \text{MIX}(\{P[w](q) \mid w \in R_f\}).$$

The idea is to cycle through all sets $U \subseteq Q$ that do not yet have a subset in $P[f](q)$, to test whether $U \in \text{MIX}(\{P[w](q) \mid w \in R_f\})$, and to add it to $P[f](q)$ if this is the case.

We do this in a bottom up fashion, starting with the singleton subsets of $U$, then testing the subsets of size 2 and so on.

Given a set $U$ that does not yet have a subset in $P[f](q)$, we test whether, for each $w \in R_f$, there is a set $W \in P[w](q)$ with $W \subseteq U$. If this is not the case, then $U \notin \text{MIX}(\{P[w](q) \mid w \in R_f\})$. If, on the other hand, this is the case, then there is a subset $U'$ of $U$ that belongs to $\text{MIX}(\{P[w](q) \mid w \in R_f\})$. In fact, we must have $U = U'$, since otherwise, we would have already added $U'$ to $P[f](q)$, and not considered $U$ for testing.

The test above can be implemented with the help of a suitable automaton. An NFA $B$ is constructed that accepts all strings $w \in \Sigma^*$ for which there is a set $W \in P[w](q)$ with $W \subseteq U$. This automaton is defined as $A_{\text{L2R}}(G)$ in Definition 3 below, but with the following modifications.

- The initial state is $\{q\}$;
- The set of accepting states is $\mathcal{P}(U)$;
- The transition function is defined with the sets $P[s](p)$ in place of $E[s](p)$, for symbols $s$ and states $p$.

That $L(B)$ is as stated above can be shown in analogy to the proof[5] of Proposition 7. Whether, for each $w \in R_f$, there is a set $W \in P[w](q)$ with $W \subseteq U$, can then be tested by checking whether $R_f \subseteq L(B)$. This latter test asks whether the language of an NFA of polynomial size is contained in the language of an NFA of exponential size. It can be translated into a nonemptiness test for an automaton of exponential size (the intersection of $B$ with the complement of the automaton for $R_f$) and is thus doable in polynomial space.

---

[5] We point out that the current proof is similar to the proof of Proposition 7 but not on the proof of Proposition 8. Rather the proof of Proposition 8 will be based on the current proof.

It remains to show that the algorithm is also correct, that is, that after its termination it holds $P[s] = E[s]$, for every $s \in \Sigma$. We do this in two steps.

We make use of the following notation. Let $P^j[s]$ denote the value of $P[s]$ after the $j$-th iteration of the WHILE loop. For a strategy $\sigma$ of JULIET and a string $u \in \Sigma^*$, we write $\text{Depth}(\sigma, u)$ for the maximum nesting depth of Call moves in any play $\Pi(\sigma, \tau, u)$. If the nesting depth is unbounded, we let $\text{Depth}(\sigma, u) = \omega$

We first show the following claim.

CLAIM 1. For every $j \geq 0$, for all symbols $\sigma \in \Sigma$ and for all $q \in Q$, $P^j[s](q)$ contains exactly the relative effects $e(\sigma, s, q)$, for all strategies $\sigma$ of JULIET with $\text{Depth}(\sigma, s) \leq j$.

We prove Claim 1 by induction on $j$.

For $j = 0$ this holds true as each $P^0[s](q) = \{\{\delta(q, s)\}\}$ is just the set with the relative effect corresponding to the strategy of JULIET that reads $s$ in its very first step.

Now let $j > 0$ and let the induction hypothesis hold for all $m < j$. We need to prove the induction step only for symbols $f \in \Gamma$ (as opposed to $s \in \Sigma$), as symbols in $\Sigma \setminus \Gamma$ only have depth-0 strategies for JULIET.

As $\{P^{j-1}[w](q) \mid w \in R_f\}$ is a finite set, there are $\ell \in \mathbb{N}$, and strings $w_1, \ldots, w_\ell \in R_f$ such that $\{P^{j-1}[w](q) \mid w \in R_f\} = \{P^{j-1}[w_i](q) \mid i \in \{1, \ldots, \ell\}\}$. For each $w$ we denote by $i(w)$ the number in $\{1, \ldots, \ell\}$ with $P^{j-1}[w](q) = P^{j-1}[w_{i(w)}](q)$. For reference, we denote the set $\{w_1, \ldots, w_\ell\}$ by $S(j, f, q)$.

Let now $e$ be a relative partial effect in $P^j[f](q)$. If $e \in P^{j-1}[f](q)$, then $e = e(\sigma, f, q)$ for some strategy $\sigma$ of JULIET with $\text{Depth}(\sigma, f) \leq j - 1$, by induction. Thus, we assume $e \in P^j[f](q) \setminus P^{j-1}[f](q)$ and thus $e$ "arrived" in $P[f](q)$ in the $j$-th iteration of the WHILE loop. Therefore, $e \in \text{MIX}(\{P^{j-1}[w](q) \mid w \in R_f\})$. Furthermore, there are relative partial effects $e_1, \ldots, e_\ell$ such that

- $e_i \in P^{j-1}[w_i](q)$, for every $i$, and
- $e = e_1 \cup \cdots \cup e_\ell$.

By induction and the correctness of $\circ$ we can conclude that, for each $i$, there is a strategy $\sigma_i$ of JULIET on $w_i$ of depth $\leq j - 1$ such that $e_i = e(\sigma_i, w_i, q)$.

The strategy $\sigma$ of depth $j$ can now be obtained as follows. In the first round, JULIET does a Call move. Then, if ROMEO chooses a string $w \in R_f$ she follows the strategy $\sigma'$ such that $e(\sigma', w, q) = e_i$, for the $i \in \{1, \ldots, \ell\}$ with $P^{j-1}[w](q) = P^{j-1}[w_i](q)$. Thus, $e(\sigma, s, q) = e_1 \cup \cdots \cup e_\ell = e$.

Conversely, let $\sigma$ be a strategy of JULIET on $f$ of depth at least 1 and at most $j$ and let $e = e(\sigma, f, q)$. The first step of JULIET, following $\sigma$, is a Call on $s$. For each $i \in \{1, \ldots, \ell\}$ let $\sigma_i$ be the strategy of JULIET that is induced by $\sigma$ on $w_i$ and let $e_i = e(\sigma_i, w_i, q)$. Now for each possible reply $w \in R_f$ of ROMEO, let $\sigma_w$ be the strategy that yields $e(\sigma_{i(w)}, w_{i(w)}, q)$ and let $e_w = e(\sigma_w, w)$. Thus,

$$e = \bigcup_{w \in R_f} e_w = e_1 \cup \cdots \cup e_\ell.$$

Clearly, each strategy $\sigma_w$, and in particular every $\sigma_{w_i}$ has a Call depth $\leq j - 1$ on $w_i$. Thus, by induction we conclude that $e_i \in P[w_i](q)$, for every $i$ and therefore $e \in \text{MIX}(\{P^{j-1}[w](q) \mid w \in R_f\})$, as required.

This concludes the proof of the Claim 1.

So far we have not ruled out that there might be a strategy $\sigma$ of JULIET with unbounded Call depth such that $e(\sigma, f, q) \notin P[s](q)$, at the end of the computation of Algorithm 1.

To bridge the gap, we use an additional game $G'$ that is obtained from $G$ by a restriction to finite rule sets as follows.

For each $f$ and each string $w \in R_f$ let $v(f, w)$ be a string of minimal length such that $v(f, w) \in R_f$ and $E[v(f, w)] = E[w]$.

Let $S$ be the union of the set of all strings of the form $v(f, w)$ with all sets[6] of the form $S(j, q, f)$ that were defined in the proof of Claim 1. Let $G' = (\Sigma, R', T)$, where, for each $f$, $R'_f = R_f \cap S$.

As all sets $S(j, q, f)$ are finite and there are only finitely many effects with respect to $G$, $S$ is a finite set and thus all sets $R'_f$ are finite as well.

CLAIM 2. For every symbol $s$, $E^G[s] = E^{G'}[s]$.

Obviously, for each $s$ and $q$ and every finite $G$-strategy $\sigma$ of JULIET, the $G'$-strategy $\sigma'$ that is induced by $\sigma$ fulfills $e^{G'}(\sigma', s, q) \subseteq e^G(\sigma'', s, q)$, simply because all plays in $G'$ are also plays in $G$.

To complete the proof of Claim 2 it thus suffices to prove the following claim.

CLAIM 3. For every string $u \in \Sigma^*$, state $q$ and finite $G'$-strategy $\sigma'$ of JULIET there is a finite $G$-strategy $\sigma$ with $e^G(\sigma, u, q) \subseteq e^{G'}(\sigma', u, q)$.

We first observe that $\text{Depth}^{G'}(\sigma', u) < \omega$ Otherwise, the strategy tree induced by $\sigma'$ on $u$ would be a finitely branching tree with arbitrarily long branches and thus would contain infinite branches by König's Lemma, contradicting finiteness.

Thus, we can show Claim 3 by induction on $\text{Depth}^{G'}(\sigma', u)$.

The case $\text{Depth}^{G'}(\sigma', u) = 0$ is simple as $G'$ and $G$ coincide as long as no Calls are made (as in the play on $u$ following $\sigma'$).

For the induction step, let $\text{Depth}^{G'}(\sigma', u) = k > 0$ and let us assume that the claim holds for smaller depths.

Let $e' = e^{G'}(\sigma', u, q)$.

We consider two cases.

The first case is that $u = sw$, for some $s \in \Sigma$ and $w \in \Sigma^*$ and $\sigma'$ plays a Read on $s$.

In this case, we can conclude that $e^{G'}(\sigma', w, p) = e'$ and that $\text{Depth}^{G'}(\sigma'_w, w) < k$, where $p = \delta(q, s)$ and $\sigma'_w$ is the strategy of JULIET on $w$ induced by $\sigma'$. Thus, by induction, there is a $G$-strategy $\sigma_w$ with $e^G(\sigma_w, w, p) \subseteq e^{G'}(\sigma_w, w, p)$. Combing $\sigma_w$ with an initial Read on $s$ yields the desired strategy $\sigma$.

The second case is that $u = fw$, for some $f \in \Sigma$ and $w \in \Sigma^*$ and $\sigma'$ plays a Call on $f$.

We define $\sigma$ as follows. For each $z \in R_f$, $v(z, f)$ is a possible answer of ROMEO in both games $G$ and $G'$. As $\text{Depth}^{G'}(\sigma', v(z, f)) < k$, induction yields a finite $G$-strategy $\sigma_{z,1}$ with $e^G(\sigma_{z.1}, v(z, f)w, q) \subseteq e^{G'}(\sigma', v(z, f)w, q) = e'$. As $E^G[v(f, w)] = E^G[v]$, there is a strategy $\sigma_z$ for JULIET with

---
[6]The sets $S(j, q, f)$ will become important after the proof of Claim 2.

$e^G(\sigma_{z.1}, v(z,f), q) = e^G(\sigma_z, z, q)$.

We define strategy $\sigma$ for the case that ROMEO replies by $z$ on $fw$ as follows. It plays according to $\sigma_z$ on $z$ and then follows the strategy induced by $\sigma_{z.1}$ on $w$. Altogether, we can conclude $e^G(\sigma, fw, q) \subseteq e'$, as required.

This completes the proof of Claim 3 and also the proof of Claim 2.

To complete the proof of the proposition it suffices to observe that, by the choice of the set $S$, the output of Algorithm 1 on input $G$ is the same as on input $G'$.

As all rule sets in $G'$ are finite, every finite strategy $\sigma'$ of JULIET contributing to $E^{G'}[s]$ are of bounded Call depth. Otherwise, the strategy tree induced by $\sigma'$ on a symbol $s$ would be a finitely branching tree with arbitrarily long branches and thus would again contain infinite branches by König's Lemma, contradicting finiteness.

Therefore, Algorithm 1 computes, on input $G$ or $G'$, all effects $E^{G'}[s]$ correctly and thus, by Claim 2, also all effects of $G$. □

### 5.3 Automata for strategies of Romeo

For the proof of Theorem 17 below, we need an NFA of exponential size for $\Sigma^* \setminus \mathrm{safe}_{L2R}(G)$. As the complementation of $A_{\mathrm{L2R}}(G)$ might yield an automaton of doubly exponential size (in $|G|$), we follow a different approach by constructing an NFA for $\Sigma^* \setminus \mathrm{safe}_{L2R}(G)$ that works analogous as $A_{\mathrm{L2R}}(G)$ but is based on strategies of ROMEO. To this end, we define dual effects and the dual automaton $\hat{A}_{\mathrm{L2R}}(G)$ next.

*Definition 4.* Let $G$ be a context-free game, $u$ a string, $q$ a state of the target automaton and $\tau$ a strategy of ROMEO. The *dual* relative effect $\hat{e}(\tau, u, q)$ is the set $\{\delta^*(q, w) | w = \mathrm{word}(u, \sigma, \tau), \sigma \in \mathrm{STRAT}_{L2R}(G)\}$.

The dual effect $\hat{E}[u]$ of $u$ maps every state $q$ to the normalized set of dual relative effects $\hat{e}(\tau, u, q)$ of $u$ for all strategies $\tau \in \mathrm{STRAT}_{\mathrm{ROMEO}}(G)$.

For the sake of clarity, we will sometimes refer to non-dual (relative) effects as *primal* (relative) effects.

The informal meaning of dual relative effects is dual to the informal meaning of primal relative effects: $\hat{e}(\tau, u, q)$ is the set of states, for which there is a strategy $\sigma$ of JULIET and a string $w \in \Sigma^*$ such that $w = \mathrm{word}(u, \sigma, \tau)$ and $\delta^*(q, w) = p$. We note that non-terminating plays do not contribute to dual effects, as for every strategy $\tau$ there is a strategy $\sigma$ of JULIET that yields a finite play (e.g., the strategy that always does Read), and thus reflecting non-terminating plays in $\hat{e}(\tau, u, q)$ would not have any consequences.

The dual effect of a string can be obtained from its primal effect via a simple operation, SMIX, very similar to the MIX operation used in previous sections. Let $\mathcal{D} = \{D_1, \ldots, D_n\}$ be a set of sets. Then

$$\mathrm{SMIX}(\mathcal{D}) = \mathrm{NORM}(\{\{d_1, \ldots, d_n\} \mid d_1 \in D_1 \wedge \cdots \wedge d_n \in D_n\}).$$

In other words, SMIX contains all sets that can be formed by selecting one element from each of the elements of $\mathcal{D}$. Notice that, while MIX takes a set of sets of sets and returns a set of sets, SMIX takes a set of sets and returns a set of sets.

LEMMA 10. *Let $u$ be a string and $q \in Q$ a state of $A(T)$. Then $\hat{E}[u](q) = \mathrm{SMIX}(E[u](q))$.*

PROOF. As both sets are normal it suffices, thanks to Lemma 3, to show that for every $\hat{e} \in \hat{E}[u](q)$ there is some $e \in \mathrm{SMIX}(E[u](q))$ such that $e \subseteq \hat{e}$, and vice versa.

Let $\hat{e} \in \hat{E}[u](q)$ and let $\tau$ be a strategy of ROMEO such that $\hat{e} = \hat{e}(\tau, u, q)$. By definition, $\hat{e} = \{\delta^*(q, w) \mid w \in \mathrm{word}(u, \sigma, \tau), \sigma \in \mathrm{STRAT}_{L2R}\}$. This means that for every $\sigma \in \mathrm{STRAT}_{L2R}$ there is a a state in $e(\sigma, u, q)$ that also belongs to $\hat{e}$. In particular, there is an element $e$ in $\mathrm{SMIX}(E[u](q))$ such that $e \subseteq \hat{e}$.

For the other direction, consider $e \in \mathrm{SMIX}(E[u](q))$. By definition of $E[u](q)$ and SMIX, for every finite strategy $\sigma$ of JULIET there is a strategy $\tau$ of ROMEO such that $\delta^*(q, w) \in e$, where $w = \mathrm{word}(u, \sigma, \tau)$.

Let $t = \mathrm{Tree}_{G,u}$ be the (full) game tree for $u$. Let $L_e$ be the set of leaves of $t$ that are labeled by configurations $(1, w, \varepsilon)$ with $\delta^*(q, w) \in e$. Let $S_e$ be the set of nodes $n$ of $t$ such that for every strategy $\sigma$ of JULIET, the subtree of the strategy tree $\mathrm{Tree}_{G,u}(\sigma)$ rooted in $n$ either has an infinite branch or a leaf in $L_e$.

The root of $t$ must belong to $S_e$. Otherwise, JULIET would have a finite strategy $\sigma$ such that no strategy of ROMEO yields a state in $e$, contradicting the above statement about $e$. Furthermore, if a node in $t$ belongs to $S_e$ and is labeled by a configuration where JULIET is to move, then all its children belong to $S_e$. If a node in $t$ belongs to $S_e$ and ROMEO is to move, then at least one of its children belongs to $S_e$. We can define a strategy $\tau$ for ROMEO that from a node in $S_e$ always selects a child node in $S_e$. In the strategy tree $\mathrm{Tree}_{G,u}(\tau)$, every node belongs to $S_e$. This immediately implies that $\hat{e}(\tau, u, q) \subseteq e$. □

From this connection it follows that composition of dual effects just works like composition of effects. In the following, the operation "∘" for dual effects is defined exactly as for effects.

LEMMA 11. *Let $u, v$ be strings. Then $\hat{E}[u] \circ \hat{E}[v] = \hat{E}[uv]$.*

PROOF. This follows from definition 4 exactly like the corresponding statement for primal effects by reversing the roles of JULIET and ROMEO in the proof of lemma 6. □

Now we are ready to define the dual automaton $\hat{A}_{L2R}(G)$ for $\Sigma^* \setminus \mathrm{safe}_{L2R}(G)$.

*Definition 5.* Let $G = (\Sigma, R, T)$ be a context-free game with a DFA $A(T) = (Q, \Sigma, \delta, q_0, F)$ for $T$. The NFA $\hat{A}_{L2R}(G) = (\hat{Q}_{L2R}, \Sigma, \hat{\delta}_{L2R}, \{q_0\}, \hat{F}_{L2R})$ is defined as follows:

- $\hat{Q}_{L2R} = \mathcal{P}(Q)$;
- $\hat{\delta}_{L2R}(X, s) = \mathrm{MIX}(\{\hat{E}[s](q) \mid q \in X\})$, for each $X \subseteq Q$ and $s \in \Sigma$;
- $\hat{F}_{L2R} = \mathcal{P}(Q \setminus F)$.

PROPOSITION 12. *Let $G = (\Sigma, R, T)$ be a context-free game with a DFA $A(T) = (Q, \Sigma, \delta, q_0, F)$ for $T$. Then $L(\hat{A}_{L2R}(G)) = \Sigma^* \setminus \mathrm{safe}_{L2R}(G)$.*

PROOF. As for $A_{L2R}(G)$, we first show that $\mathrm{NORM}(\hat{\delta}^*_{L2R}(q_0, u)) = \hat{E}[u](q_0)$, by induction on $|u|$.

For $u = \epsilon$ we have

$$\mathrm{NORM}(\hat{\delta}^*_{L2R}(\{q_0\}, \epsilon)) = \{\{q_0\}\} = \hat{E}[\epsilon](q_0).$$

For $u = vs$ we get

$$
\begin{aligned}
N(\hat{\delta}^*_{\text{L2R}}(\{q_0\}, vs)) &= N(\bigcup_{X \in \hat{\delta}^*_{\text{L2R}}(\{q_0\},v)} \hat{\delta}_{\text{L2R}}(X, s)) \\
&= N(\bigcup_{X \in \hat{E}[v](q_0)} \hat{\delta}_{\text{L2R}}(X, s)) \\
&= N(\bigcup_{X \in \hat{E}[v](q_0)} \text{Mix}(\{\hat{E}[s](q) \mid q \in X\})) \\
&= (\hat{E}[v] \circ \hat{E}[s])(q_0) \\
&= \hat{E}[v](q_0).
\end{aligned}
$$

We can conclude as follows that ROMEO has a L2R winning strategy on $u$ if and only if $\hat{A}_{\text{L2R}}(G)$ accepts $u$.

$$
\begin{aligned}
u \in \Sigma^* \setminus \text{safe}_{L2R}(G) &\Leftrightarrow \exists e \in \hat{E}[u](q_0) : e \cap F = \emptyset \\
&\Leftrightarrow \hat{E}[u](q_0) \cap \mathcal{P}(Q \setminus F) \neq \emptyset \\
&\Leftrightarrow N(\hat{\delta}^*_{\text{L2R}}(\{q_0\}, u)) \cap \mathcal{P}(Q \setminus F) \neq \emptyset \\
&\Leftrightarrow \hat{\delta}^*_{\text{L2R}}(\{q_0\}, u) \cap \mathcal{P}(Q \setminus F) \neq \emptyset
\end{aligned}
$$

□

PROPOSITION 13. *There is an algorithm that computes in exponential time the NFA $\hat{A}_{L2R}(G)$ for each context-free game $G = (\Sigma, R, T)$, provided that $T$ is represented by a DFA $A(T) = (Q, \Sigma, \delta, q_0, F)$ and the sets $R_f$ are represented by DFAs, NFAs or regular expressions.*

PROOF. Similar to the algorithm computing $A_{\text{L2R}}(G)$, this algorithm first computes in exponential time the effects $E[s]$, for every symbol $s \in \Sigma$. From these, it computes $\hat{E}[s]$, for every $s \in \Sigma$, via $\hat{E}[s](q) = \text{SMix}(E[s](q))$, for every $q \in Q$. Each computation of a set $\text{SMix}(E[s](q))$ can be done in exponential time, similarly as for line 5 of Algorithm 1. To this end, one can test, for every set $X \subseteq Q$, whether it can be obtained by picking elements from the sets in $E[s](q)$. The sets $\text{Mix}(\{\hat{E}[s](q) \mid q \in X\})$ can be computed in exponential time as well in a straightforward fashion. □

## 6. UPPER BOUNDS

In this section, we prove the upper bounds results of our Main Theorem 1. The problem L2RALL is decidable and can actually be decided in exponential space. If all rule sets are finite and given in the input explicitly, then the problem can be decided in exponential time.

Before we describe the algorithm for L2RALL, we state two auxilliary results that allow us to consider only finite subsets of each replacement languages.

For any string $w$, let $F(w) = \{q \in Q \mid E[w](q) \cap \mathcal{P}(F) \neq \emptyset\}$ be the set of states from which JULIET has a winning strategy on $w$.

For a state $q$ and a set $S$ of states let $A^{q,S}_{\text{L2R}}$ denote the automaton that is obtained from $A_{\text{L2R}}(G)$ by chosing $q$ as initial state and $\mathcal{P}(S)$ as set of accepting states.

LEMMA 14. *For every state $q$ and $w \in \Sigma^*$ the automaton $A^{q,F(w)}_{L2R}$ accepts exactly the strings $v$ such that there is a winning strategy for JULIET on $vw$ starting at state $q$ in $A(T)$.*

The proof is similar to the proof of Proposition 7.

For a state $q \in Q$ let $G_q$ denote the game obtained from $G$ by chosing the state $q$ as initial state of the target automaton.

LEMMA 15. *Let $q \in Q$, $w \in \Sigma^*$ and $f \in \Gamma$. If there is a string $v \in R_f$ such that $vw \in \text{safe}_{L2R}(G_q)$ then there is a string $v'$ of length at most $|Q_f| \cdot 2^{|Q|}$ such that $v'w \in \text{safe}_{L2R}(G_q)$.*

PROOF. This follows from Lemma 14 and a standard pumping argument for the product automaton $B$ combining $\hat{A}_{\text{L2R}}$ and $A(R_f)$: For any two states $(X_1, p_1), (X_2, p_2) \in \hat{Q}_{\text{L2R}} \times Q_f$ there is a string $v$ with $\delta_B((X_1, p_1), v) = (X_2, p_2)$ if and only if there is such a string $v'$ of length at most $|\hat{Q}_{\text{L2R}} \times Q_f| = |Q_f| \cdot 2^{|Q|}$. □

THEOREM 16. L2RALL ∈ EXPSPACE

PROOF. We give a nondeterministic exponential-space algorithm $\mathcal{A}$ deciding $\overline{\text{L2RALL}}$, the complement of L2RALL. This yields the result since EXPSPACE is closed under complement and NEXPSPACE = EXPSPACE thanks to Savitch's Theorem [9].

The idea is that $\mathcal{A}$ guesses a symbol $f \in \Gamma$ and strings $u, w$ such that $ufw \in \text{safe}_{L2R+}(G) \setminus \text{safe}_{L2R}(G)$ is a witness string on which JULIET plays Call in the first pass on $ufw$. Thanks to Lemma 15, $\mathcal{A}$ only needs to verify that, for all replacement strings $v \in R_f$ of length at most $|Q_f| \cdot 2^{|Q|}$, it holds that $uvw \in \text{safe}_{L2R}(G)$. A short summary of $\mathcal{A}$ is given as Algorithm 2.

---

**Algorithm 2** Test for $G \in$ L2RALL

1: Guess $f \in \Gamma$ and a dual relative effect $\hat{e}_{uf}$
2: **while** Guessing a string $u$ in a streaming fashion **do**
3:   Use $A_{\text{L2R}}(G)$ to compute the set $U = E[u](q_0)$ non-deterministically
4:   Use $\hat{A}_{\text{L2R}}(G)$ to nondeterministically verify $\hat{e}_{uf} \in \hat{E}[uf](q_0)$
5: Guess a string $w$ and compute $F(w)$ by simulating $A_{\text{L2R}}(G)$ backwards
6: **if** $\hat{e}_{uf} \cap F(w) = \emptyset$ **then**
7:   // $ufw \notin \text{safe}_{L2R}(G)$
8:   **for all** $v \in R_f$ with $|v| \leq |Q_f| \cdot 2^{|Q|}$ **do**
9:     Guess a set $U_v \in U$
10:    **for all** $q \in U_v$ **do**
11:      Simulate $A^{q,F(w)}_{\text{L2R}}$ on input $v$
12:      **if** $A^{q,F(w)}_{\text{L2R}}$ accepts $v$ **then**
13:        // $uvf \in \text{safe}_{L2R}(G)$
14:      **else**
15:        Reject
   Accept
16: Reject

---

The main challenge is that the string $u$ may in general be of doubly exponential length and therefore cannot be stored.

Therefore, to compute the sets $U = \{U_1, U_2, \ldots, U_n\}$, $\mathcal{A}$ guesses $u$ in a streaming fashion, one symbol at a time. It simulates $A_{\text{L2R}}$ on $u$ and computes $E[u](q_0)$ online. This can be done in exponential space by storing the set $E[u](q_0) \in \mathcal{P}(\mathcal{P}(Q))$. At the same time, having guessed a dual relative effect $\hat{e}_{uf}$, it guesses a run of $\hat{A}_{\text{L2R}}$ on $uf$, effectively verifying that there is a strategy corresponding to this relative effect.

Afterwards, to compute $F(w) \in \mathcal{P}(Q)$, $\mathcal{A}$ guesses a string $w$, and incrementally computes a set $F(w) \subseteq Q$ of states from which JULIET can win the game as defined in Lemma 14. The set $F(sw')$ can be computed from the set $F(w')$ by checking, for each $q \in Q$, whether $A_{\text{L2R}}^{q,F(w')}$ accepts $s$. As there are only exponentially many subsets of $Q$ it is not hard to prove by a standard pumping argument that $w$ can be chosen of exponential size and that its computation can be actually carried out in polynomial space. With $F(\epsilon) = F$ the correctness of this incremental computation follows by a simple induction argument.

The algorithm then checks whether $\hat{e}_{uf}$ contains a state from $F(w)$. If it does not, we know that $ufw \notin \text{safe}_{L2R}(G)$. If it does, $\mathcal{A}$ immediately rejects.

Finally, $\mathcal{A}$ checks for all strings $v \in R_f$ of length at most $|Q_f| \cdot 2^{|Q|}$, if $uvw \in \text{safe}_{L2R}(G)$. This can be done by (1) cycling through all strings $v$ of this length, (2) checking if $v \in R_f$ by simulating $A(R_f)$ on $v$ and (3) in case $A(R_f)$ accepts $v$, guessing a set $U_i \in U$ and testing whether for every $q \in U_i$ there is a relative effect $e \in E[v](q)$ such that $e \subseteq F(w)$.

To perform test (3), $\mathcal{A}$ simulates, for each $q \in U_v$, a run of $A_{\text{L2R}}^{q,F(w)}$ on $v$. This can be done in PSPACE. If all runs succeed, $\mathcal{A}$ concludes that $uvw \in \text{safe}_{L2R}(G)$, otherwise it rejects.

Altogether, $\mathcal{A}$ only requires exponential space; it remains to show that $\mathcal{A}$ accepts iff $\text{safe}_{L2R+}(G) \setminus \text{safe}_{L2R}(G) \neq \emptyset$.

If $\mathcal{A}$ accepts, then there exists a string $ufw$ such that (a) $ufw \notin \text{safe}_{L2R}(G)$ (this follows directly from Lemma 14) and (b) for all $v \in R_f$ of length at most $|Q_f| \cdot 2^{|Q|}$ there exists a set $U_v \in E[u](q_0)$ such that $v$ is accepted by $A_{\text{L2R}}^{q,F(w)}$ for all $q \in U_v$.

With Lemmas 14 and 15, it follows from (b) that for every $v \in R_f$ there is a strategy $\sigma_v$ of JULIET on $u$ such that for all states $q \in e(u, \sigma_v, q_0)$, JULIET has a winning strategy on $vw$ starting at $q$.

This yields a winning L2R$^+$strategy for JULIET on $ufw$: In the first pass, JULIET calls $f$. On the second pass, depending on ROMEO's choice of $v$, JULIET plays according to $\sigma_v$ on $u$ and is guaranteed to reach a state starting from which she has a winning strategy on $vw$.

For the "only if" part, assume $\text{safe}_{L2R+}(G) \setminus \text{safe}_{L2R}(G) \neq \emptyset$ holds. Then there exists a word on which JULIET has a winning L2R$^+$strategy, but no winning L2R strategy. This word must be of the form $ufw$ with $f$ being the symbol JULIET calls on her first pass for some winning L2R$^+$strategy $\sigma$. In lines 1 through 4, $\mathcal{A}$ guesses this word.

Since JULIET has no winning L2R strategy on $ufw$, ROMEO must have a strategy $\tau$ on $uf$ such that $\hat{e}(uf, \tau, q_0) \cap F(w) = \emptyset$. Since this dual relative effect can be guessed by $\mathcal{A}$, the test on line 6 can be passed.

Let $\sigma_v$ be JULIET's strategy on $u$ in case ROMEO replaces $f$ by $v \in R_f$ and $U_v = e(u, \sigma_v, q_0) \in E[u](q_0)$. Since $\sigma$ is winning on $uvw$, JULIET has a winning strategy on $vw$ starting at $q$ for any $q \in U_v$. Using Lemma 14, this means that for any $v \in R_f$, $\mathcal{A}$ can guess a set $U_v \in E[u](q_0) = U$ on line 9 such that all $A_{\text{L2R}}^{q,F(w)}$ accepts $v$ for $q \in U_v$. This condition is checked in lines 10 through 15, and since it is fulfilled for all $v \in R_f$, $\mathcal{A}$ accepts. □

For games $G$ with finite replacement languages, this algorithm can be modified to run in exponential time in $|G|$.

THEOREM 17. L2RALL $\in$ EXPTIME *for games with finite replacement languages, given explicitly as part of the input.*

PROOF. We are going to modify Algorithm 2 such that it runs in exponential time. This works because the only NFAs of doubly exponential size that Algorithm 2 uses, can be replaced by NFAs of exponential size, if the replacement sets $R_f$ are finite and explicitly given in the input.

Algorithm 2 uses nondeterminism for two kinds of purposes: for guessing effects and other sets and for guessing strings. The latter can be delegated to standard polynomial space non-emptiness tests for exponential size automata, while the former can be done by cycling through all possible candidates (as there are always only exponentially many).

To this end, the algorithm $\mathcal{A}'$ contains an outer loop over all $f \in \Gamma$, sets $W \subseteq Q$ and vectors of sets $U_1, \ldots, U_{|R_f|} \in \mathcal{P}(Q)$. Inside this loop, similar to algorithm 2, $\mathcal{A}'$ checks if there are strings $u$ and $w$ such that $U_1, \ldots, U_{|R_f|} \in E[u](q_0)$ and $W = F(w)$; then, all $\mathcal{A}'$ needs to do is check for all $i = 1, \ldots, |R_f|$ whether $\delta_{\text{L2R}}^*(U_i, v_i) \cap \mathcal{P}(F(w)) \neq \emptyset$ (with $R_f = \{v_1, \ldots, v_{|R_f|}\}$) and $\hat{A}_{\text{L2R}}$ accepts $ufw$.

To verify the existence of a string $u$ with $U_1, \ldots, U_{|R_f|} \in E[u](q_0)$, $\mathcal{A}'$ computes the product automaton of $|R_f|$ copies of $A_{\text{L2R}}$ and checks whether the product state $(U_1, \ldots, U_{|R_f|})$ is reachable inpolynomial space (and thus exponential time).

To find a string $w$ with $W = F(w)$, $\mathcal{A}'$ computes the product automaton with one copy of $A_{\text{L2R}}{}^{q,F}$, for each $q \in W$; again, the verification of the existence of $w$ is by a non-emptiness test.

Finally, $\mathcal{A}$ runs one copy of $A_{\text{L2R}}$ with starting state $U_i$ and final state set $\mathcal{P}(W)$ on $v_i$ for each $i = 1, \ldots, |R_f|$ and runs $\hat{A}_{\text{L2R}}$ on $ufw$; if all copies of $A_{\text{L2R}}$ and $\hat{A}_{\text{L2R}}$ accept, $\mathcal{A}$ accepts, since a separating string $ufw$ has been found. The correctness of this algorithm follows similar to the proof of theorem 16, and since it loops an exponential number of times and takes no more than exponential time in each iteration, $\mathcal{A}'$ is an EXPTIME algorithm deciding L2RALL for games with finite replacement languages, given explicitly as part of the input. □

## 7. LOWER BOUNDS

In this section, we prove the hardness results of Theorem 1. More precisely, we show that L2RALL is EXPSPACE-hard in general and EXPTIME-hard for games with finite replacement sets.

PROPOSITION 18. L2RALL *is hard for* EXPSPACE.

PROOF. We prove the EXPSPACE hardness of L2RALL by reduction from the EXPONENTIAL WIDTH CORRIDOR TILING problem. In this problem, we are given a set $U = \{u_1, \ldots, u_k\}$ of tiles with a designated *initial tile* $u_I \in U$ and *final tile* $u_F \in U$. There are also two relations $H, V \subseteq U \times U$. These are the *horizontal* and *vertical* constraints, respectively. A tile $u_j$ is only allowed to the right of a tile $u_i$ if $(u_i, u_j) \in H$ and only allowed on top of $u_i$ if $(u_i, u_j) \in V$. We are also given a number $n$ in unary notation.

Formally, a corridor tiling of width $\ell$ is a mapping $t : \{0, \ldots, \ell - 1\} \times \{0, \ldots, m\} \to U$, for some $m$. A tiling $t$ is *valid* if

- $t(0, 0) = u_I$,
- $t(\ell - 1, m) = u_F$,

- for every $i \in \{0, \ldots, \ell-2\}$ and $j \in \{0, \ldots, m\}$, $(t(i,j), t(i+1,j)) \in H$, and
- for every $i \in \{0, \ldots, \ell-1\}$ and $j \in \{0, \ldots, m-1\}$, $(t(i,j), t(i,j+1)) \in V$.

EXPONENTIAL WIDTH CORRIDOR TILING asks whether an instance $\mathcal{I} = (U, u_I, u_F, V, H, n)$ has a valid corridor tiling of width $2^n$. This problem is well known to be EXPSPACE-complete; see, e.g., [4, 10].

Given an input instance $\mathcal{I} = (U, u_I, u_F, V, H, n)$ for EXPONENTIAL WIDTH CORRIDOR TILING, we construct from $\mathcal{I}$ a context-free game $G = (\Sigma, R, T)$ such that there exists a valid corridor tiling for $\mathcal{I}$ if and only if there is a string for which JULIET has a winning L2R$^+$strategy but no L2R strategy in $G$. The claim then follows from this reduction by Lemma 4.

The rough idea of the construction of $G$ is as follows. Let $2^n$ be the target width for a tiling. Tilings are encoded by strings of the form $v = ((uc)^*\#)^*$, where $u$ is a tile and $c$ a 0-1-string of length $n$ that should encode the column number of the position of $u$. A sequence $(uc)^*$ encodes a row of a tiling and rows are separated by $\#$. For each column number $i \in \{0, 1, \ldots, 2^n - 1\}$, we denote by $c(i)$ the encoding of $i$ as a binary string of length $n$ over $\{0, 1\}$.

We will construct $G$ in such a way that the strings in safe$_{\text{L2R+}}(G) \setminus$ safe$_{L2R}(G)$ are of the form $gvf$, where $v$ is the encoding of a correct tiling.

The main task of JULIET in the game is to show that the middle part $v$ of the input string indeed represents a correct tiling, while ROMEO tries to disprove her. For this, we utilize a *protest technique* [8], in which we force JULIET to call potentially inconsistent symbols in the input, allowing ROMEO to flag constraint violations. The additional symbols $f$ and $g$ are primarily meant to ensure that JULIET needs a L2R$^+$strategy to win; $f$ will also be needed to identify violations of vertical constraints, as we shall describe later.

Before giving $G$ in formal detail, we shall first describe the ways in which a string $v$ of the form $(U0^n(U\{0,1\}^n)^*U1^n\#)^*$ may fail to encode a valid tiling. After that, we examine how to deal with these types of violations.

- *Horizontal error:* $v$ violates the horizontal constraints, i.e. $v$ contains a substring of the form $u\{0,1\}^n u'$ with $(u, u') \notin H$;
- *Constant error:* The first (last) symbol from $U$ in $v$ is not $u_I$ ($u_F$);
- *Increment error:* Two subsequent column number encodings are inconsistent, i.e. $v$ contains a substring of the form $c(i)Uc(j)$ with $j \neq i + 1$;
- *Vertical error:* $v$ violates the vertical constraints, i.e. $v$ contains a substring of the form $uc(i)(U \cup \{0,1\})^*\#(U \cup \{0,1\})^*u'c(i)$ with $(u, u') \notin V$ for some $i \leq 2^n - 1$.

We will construct $G$ such that ROMEO can win without any effort on inputs with horizontal or constant errors and by pinpointing positions with increment or vertical errors otherwise. Horizontal and constant errors can be basically tested by the target DFA, so we merely need to make certain that strings with these kinds of errors can never be rewritten into the target language.

In the main part of the game, during the second pass, JULIET calls all positions of tiling symbols and gives ROMEO the possibility to mark a violating position. If $v$ contains an increment error at some position, ROMEO can prove this with a simple subgame. Verifying vertical errors is slightly more complicated. To this end, JULIET has to allow ROMEO to add an $n$-digit number $c_f$ to the end of $v$ in the single move of the first pass. ROMEO should pick $c_f$ as the encoding of the number of a column in which a vertical error occurs. In the main part, ROMEO can then indicate the positions of the two tiles of that error and in a subgame it is verified that they are actually in the same column (with number $c_f$) on consecutive rows.

We force JULIET to call all positions of tiling symbols by introducing into the alphabet a disjoint copy $\hat{U}$ of $U$, the set of *marked tiles*, as well as a *protest symbol* @. The idea is that for as long as JULIET keeps calling tile positions in order, ROMEO replaces those tiles with their corresponding marked tiles, but as soon as JULIET skips a tile, ROMEO protests by returning @ the next time JULIET plays a call move. By including only appropriate strings in the target language, we make sure that ROMEO wins on strings on which JULIET has tried to "cheat" by skipping a tile and ROMEO has rightfully protested, and that ROMEO loses on strings on which he protests without just cause.

Increment errors are dealt with in a similar manner by use of a *number protest symbol* @$_N$. As soon as JULIET calls the tile position immediately before the violating substring $c(i)Uc(j)$, ROMEO returns @$_N$, signifying that JULIET now has to call each of the $n$ bits to the right of @$_N$ in turn until ROMEO returns a *flag bit* $0_N$ or $1_N$ to pinpoint a position in $c(i)$ that witnesses $j \neq i+1$. (The correctness of this flagging procedure will follow from Lemma 20 below.) Similarly as for tiles, we use additional *marked bits* $\hat{0}, \hat{1}$ and the protest symbol @ to force JULIET to call each position of $c(i)$.

Finally, to handle vertical errors, we add another disjoint copy $U^V$ of $U$, called *flagged tiles* to the alphabet. As described above, after giving the encoding $c_f$ of a column where vertical constraints are violated, ROMEO replaces two tiles involved in this violation by their corresponding flagged tiles. Again, we need to make sure via the target language that ROMEO always wins on rewritten strings with correctly flagged vertical errors and loses on strings with incorrect claims of vertical errors.

Now, we can begin constructing the game $G = (\Sigma, R, F)$ from the tiling input $\mathcal{I} = (U, u_I, u_F, V, H, n)$ according to the ideas laid out above

The alphabet $\Sigma$ consists of the union of $U = \{u_1, \ldots, u_k\}$ with two disjoint copies of $U$, called $\hat{U}$, and $U^V$ with symbols $\hat{u}_1, \ldots, \hat{u}_k$ and $u_1^V, \ldots, u_k^V$, respectively and the additional symbols $0, 1, \hat{0}, \hat{1}, 0_N, 1_N, \#, @, @_N, f, g, g', h, h'$. To make the definition of $T$ somewhat more concise, we also give names to several subsets of $\Sigma$:

- the set of *base symbols* $\Sigma_B = U \cup \{0, 1, \#\}$;
- the set of *processed symbols* $\Sigma_P = \hat{U} \cup \{0, 1, \#\}$;
- the set of *extended bits* $\hat{B}_N = \{0, 1, \hat{0}, \hat{1}, 0_N, 1_N\}$.

The set $R$ consists of the following replacement rules:[7]

$$\begin{aligned} g &\to g' \\ u_1 &\to \hat{u}_1 \mid u_1^V \mid @ \mid @_N \\ &\vdots \\ u_k &\to \hat{u}_k \mid u_k^V \mid @ \mid @_N \\ 0 &\to \hat{0} \mid 0_N \mid @ \\ 1 &\to \hat{1} \mid 1_N \mid @ \\ f &\to \{0,1\}^n\{h,h'\} \mid @ \end{aligned}$$

The target language $T$ is the union of several languages described below. To improve readability, we give these languages as regular expressions, but it is easy to verify that a DFA of polynomial size in $|\mathcal{I}|$ accepts each of them. In these regular expressions, we use the abbreviations $(\alpha)_{=n}$ and $(\alpha)_{<n}$ to denote strings of length exactly $n$ (respectively less than $n$) described by the regular expression $\alpha$. Again, it is easy to verify that the size of a DFA for $(\alpha)_{=n}$ and $(\alpha)_{<n}$ is at most $n$ times the size of a DFA for $\alpha$. We shall also use the following common shorthand notations: $\alpha^?$ for $\alpha+\epsilon$; $\alpha^n$ for the $n$-fold concatenation of $\alpha$ with itself; and $\alpha^{n+}$ for $\alpha^n \alpha^*$.

As a further shorthand notation,

- let $W$ denote the set of all *well-formed tiling encodings* described by $(U0^n(U\{0,1\}^n)^*U1^n\#)^*$;

- let $H$ denote the set of *horizontally correct encodings*, which is obtained by taking the complement of the language described by $\bigcup_{(u,u')\notin H} \Sigma_B^* u\{0,1\}^* u' \Sigma_B^*$; and

- let $C$ denote the set of all *tiling candidates* which are strings that belong to $W$ (i.e. have column numbers of length $n$ that start at 0 and end at $2^n - 1$ in each line), are also in $H$ (i.e. satisfy the horizontal constraints) and have $u_I$ as their first and $u_F$ as their last tile.

Let $\hat{W}, \hat{H}, \hat{C}$ be defined as $W, H$ and $C$, but with $\hat{U}$ in place of $U$.

It is straightforward to construct a polynomial-size DFA for the set $C$.

With these notations, we define the target language $T$ as the union of the following languages (a short intuitive description follows below):

(1) $gC@$

(2) $g\hat{C}\{0,1\}^n h + g'\hat{C}\{0,1\}^n h'$

(3) $(g+g')(\Sigma_P + U^V)^* @\Sigma_B^*(h+h')$

(4) $(g+g')(\Sigma_P + U^V)^* @_N F_I \Sigma_B^*(h+h')$, where $F_I$ is the

---

[7] Note that, even though all replacement languages are finite, Theorem 17 does not apply here, since the $2^{n+1}$ replacement strings of the replacement rule $f \to \{0,1\}^n\{h,h'\}$ are not given explicitly but by a DFA of size $O(n)$.

language described by

$$(\hat{0}+\hat{1})^* 0_N (1^*U(0+1)^*)_{=n} 10^*(U+\#)+ \tag{i}$$

$$(\hat{0}+\hat{1})^* 1_N (1^*U(0+1)^*)_{=n} 00^*(U+\#)+ \tag{ii}$$

$$(\hat{0}+\hat{1})^* 0_N (0(0+1)^*U(0+1)^*)_{=n} 0(0+1)^*(U+\#)+ \tag{iii}$$

$$(\hat{0}+\hat{1})^* 0_N ((0+1)^*U(0+1)^*)_{=n} 01(0+1)^*(U+\#)+ \tag{iv}$$

$$(\hat{0}+\hat{1})^* 1_N (0(0+1)^*U(0+1)^*)_{=n} 1(0+1)^*(U+\#)+ \tag{v}$$

$$(\hat{0}+\hat{1})^* 1_N ((0+1)^*U(0+1)^*)_{=n} 11(0+1)^*(U+\#)+ \tag{vi}$$

$$(\hat{0}+\hat{1})^* (0_N + 1_N)(0+1)^* \#+ \tag{vii}$$

$$(\hat{0}+\hat{1})^* \hat{0}(\hat{0}+\hat{1})^* U + \hat{1}^* \# \tag{viii}$$

(5) $(g+g')\Sigma_P^*(U^V\Sigma_P^*)^?(U^V(0+1)^* + @_N(\hat{0}+\hat{1})^*)@\Sigma_B^*(h+h')$

(6) $(g+g')(\Sigma_P^* U^V \hat{B}_N^*)^3 \Sigma_B^*(h+h')$

(7) $(g+g')\Sigma_P^* U^V \hat{B}_N^*(\hat{U}+\#)\Sigma_P^*(h+h')$

(8) $\bigcup_{(u,v)\in V}(g+g')\Sigma_P^* u^V (\Sigma_P + \hat{B}_N)^* v^V \Sigma_B^*(h+h')$

(9) $(g+g')\Sigma_P^* U^V (\Sigma_P\setminus\{\#\})^*((\#(\Sigma_P\setminus\{\#\})^*)^{2+})^? U^V \Sigma_B^*(h+h')$

(10)

$$\begin{aligned} &\left\{ (g+g')\Sigma_P^* U^V(0+1)^*(\hat{0}+0_n) \left[((\Sigma_P\setminus\{\#\})^n + \right.\right. \\ &\left.\left. (\Sigma_P^* \# \Sigma_P^*)_{=n+1})(0+1)\right]^* 1(0+1)_{<n-1}(h+h') \right\} + \\ &\left\{ (g+g')\Sigma_P^* U^V(0+1)^*(\hat{1}+1_n) \left[((\Sigma_P\setminus\{\#\})^n + \right.\right. \\ &\left.\left. (\Sigma_P^* \# \Sigma_P^*)_{=n+1})(0+1)\right]^* 0(0+1)_{<n-1}(h+h') \right\} \end{aligned}$$

Recall that the purpose of this construction is to allow JULIET to win in $G$ with a L2R[+] but not a L2R strategy on exactly the strings $gvf$ where $v$ encodes a valid tiling for $\mathcal{I}$. To this end, we force her to perform her initial first-pass call on the final $f$, allowing ROMEO to fix a column number $c_f \in \{0,1\}^n$ for finding vertical errors. To ensure that JULIET needs a L2R[+] strategy to win on $gvf$, ROMEO may append either $h$ or $h'$ and we expect the first symbol of words in $T$ to match their last symbol; therefore, part (2) of the above definition corresponds to the case where JULIET and ROMEO play according to the intuitive rules described above, $v$ encodes a valid tiling and ROMEO never tries to protest. All the other parts of the target language serve only to prevent unjustified protest by ROMEO and to make sure that he loses immediately if he tries to "cheat" by claiming an inconsistency where there is none.

Parts (1), (3) and (5) of the target language address unjustified protests against the sequence in which JULIET calls input symbols, forcing ROMEO to reserve the protest symbol @ for cases when JULIET skips an input symbol she is supposed to call. Part (4) prevents unjustified claims of an incremental error, with sub-expressions roughly corresponding to the different cases of Lemma 20. Parts (6) to (10) deal with attempts to wrongly claim a vertical error: Flagging too many tiles (6) or not enough tiles (7), flagging two tiles not violating the vertical constraints (8), flagging tiles in non-subsequent rows (9) and flagging tiles not in the column determined by $c_f$ (10).

We now state the main ingredient of our proof:

LEMMA 19. *Let $c_f \in \{0,1\}^n$ be the binary encoding of $n_f \in \{0, 1, \ldots, 2^n - 1\}$ and let $v$ be a string of the form $(U0^n(U\{0,1\}^n)^*U1^n\#)^*$. Then the following statements are equivalent:*

*(a) $v$ is a tiling candidate for $\mathcal{I}$ that has no vertical errors in column $n_f$ and no incremental errors*

*(b) $gvc_fh \in \text{safe}_{L2R}(G)$*

*(c) $g'vc_fh' \in \text{safe}_{L2R}(G)$*

PROOF OF LEMMA 19. We shall only prove $(a) \Leftrightarrow (b)$, the proof of $(a) \Leftrightarrow (c)$ is analogous.

"$(a) \Rightarrow (b)$":

We shall describe a winning L2R strategy on $gvc_fh$ for JULIET.

During her left-to-right pass, JULIET proceeds to call every symbol from $U$ for as long as ROMEO returns only symbols from $\hat{U}$. If she reaches the end of the string this way, she wins due to part (2) of $T$. If, at some point, ROMEO decides to raise a false protest that JULIET is not calling all symbols from $U$ in sequence (i.e. returns @), JULIET wins by (3).

If ROMEO raises a false protest about an incorrectly encoded index (i.e. returns $@_N$ to one of JULIET's calls), JULIET proceeds to call all the bits to the right of $@_N$ in a left-to-right manner. This can result in the following:

- ROMEO returns @ at some point: In this case, JULIET wins by (5);

- ROMEO returns $\hat{0}$ or $\hat{1}$ for every bit of the index string to the right of $@_N$: In this case, JULIET wins by line (viii) of (4).

- ROMEO returns $0_N$ or $1_N$ at some point: In this case, JULIET wins by one of the other parts of (4) and Lemma 20 as we shall explain below.

Let $c(i)$ be the column index to the right of $@_N$ and $k$ the index of the bit in $c(i)$ for which ROMEO returned a flagged bit.

If $c(i)$ is followed by $\#$, JULIET wins by line (vii) of (4). Therefore assume that $@_N$ is followed by a string of the form $c(i)Uc(j)$. Since $v$ contains no increment errors, the contraposition of parts (a), (b) and (c) of Lemma 20 holds.

If $c(i)_k \neq c(j)_k$, then JULIET wins by (i) or (ii) due to part (a) of Lemma 20; if $c(i)_k = c(j)_k$, JULIET wins by one of (iii)-(vi) due to part (b) of Lemma 20. Since there are no other cases, JULIET can always play to win if ROMEO tries to protest an incremental error.

It remains to be explained how JULIET plays if ROMEO returns a symbol from $U^V$ to one of JULIET's calls on the symbols from $U$ (i.e. protests falsely about a vertical error). If this happens and the $n$-bit string $c(i)$ to the right of ROMEO's protest does not equal $c_f$, JULIET calls a bit in $c(i)$ on which $c(i)$ and $c_f$ differ. If ROMEO answers this call with @, he loses by (5), if he answers with $\hat{0}$ or $0_N$ ($\hat{1}$ or $1_N$), he loses by the first (second) term of (10).

If $c_i$ matches $c_f$, JULIET continues calling all occurrences of symbols from $U$. If ROMEO raises no further protest (or protests with @ or $@_N$, as above), he loses by (7) (or (3), (5) or (4) as above). If he flags another tile by returning a symbol from $U^V$ with column index string $c(j)$, he loses as described above if $c(j) \neq c_f$. If $c(j) = c_f$, again, JULIET continues calling all symbols from $U$. Should ROMEO then return anything but a symbol from $\hat{U}$ to any of JULIET's calls, she wins as described above by reaching a word in (3), (4), (5) or (6).

Finally, if the word reached after JULIET has called all the occurrences of symbols from $U$ contains exactly two tiles $u_1^V, u_2^V \in U^V$, by the above strategy, their column index strings $c(i)$ and $c(j)$ both have to match $c_f$. Therefore $i = j$, which means that $u_1$ and $u_2$ have to be in the same column of the tiling candidate encoded in $v$. Thus, $u_1$ and $u_2$ are either in non-subsequent lines (in which case JULIET wins by (9)) or conform to the vertical constraints (in which case JULIET wins by (8)).

As the above cases cover all possible counter-strategies of ROMEO, the above is a winning L2R$^+$ strategy for JULIET, and it follows that $gvf \in \text{safe}_{L2R+}(G)$

"$(b) \Rightarrow (a)$":

Assume that JULIET has a winning L2R strategy $\sigma$ on $gvc_fh$.

On her pass through $v$ according to $\sigma$, JULIET has to call every occurrence of a symbol from $U$ for as long as ROMEO keeps returning symbols from $\hat{U}$, because if she were to skip a symbol from $U$, ROMEO could respond to her next call with @ and she would lose. Leaving any symbols from $U$ uncalled to deny ROMEO the option of protesting with @ is not an option, either, because there are no words in $T$ containing symbols from $U$ without also containing symbols from $U^V \cup \{@, @_N\}$. Also, for as long as ROMEO keeps returning symbols from $\hat{U}$, JULIET may not call any symbol not in $U$ (i.e. 0 or 1), since all words containing $\hat{0}, \hat{1}, 0_N$ or $1_N$ also have to contain a symbol from $U^V \cup \{@_N\}$ further to the left.

As JULIET has to win the game if ROMEO only returns symbols from $\hat{U}$, it follows that $v \in C$, since the only winning condition not involving any protest symbols is (2). Therefore $v$ encodes a tiling candidate, i.e. contains no horizontal or constant errors.

Now assume for the sake of contradiction that $v$ contains an increment error, i.e. a substring of the form $uc(i)u'c(j)$ with $u, u' \in U, c(i), c(j) \in \{0,1\}^n$ and $i + 1 \neq j$.

In this case, ROMEO has a winning strategy in which he responds with $@_N$ as soon as JULIET calls $u$. After this move, JULIET is forced to call all bits from $\{0,1\}$ to the right of $u$ until ROMEO responds with $0_N$, $1_N$ or @ or the next symbol not in $\{0,1\}$ is reached. This is because every string in $T$ that contains $@_N$ requires one of these characters to its right, separated from $@_N$ by only characters from $\{\hat{0}, \hat{1}\}$.

Since $uc(i)u'c(j)$ is an increment error, one of the three conditions of the conclusion of Lemma 20 holds. If (c) holds, then $c(i) = 1^n$ and JULIET may win by replacing each bit of $c(i)$ by $\hat{1}$, since line (viii) of part (4) of the target language only allows $\hat{1}^n$ to be followed by $\#$, not $u'$. If (a) or (b) hold, then there exists a position $k$ in $c(i)$ such that ROMEO may return a flagged bit $0_N$ or $1_N$ on JULIET's call on $c(i)_k$, (a) prevents her from winning according to lines (i) or (ii) of (4) and (b) prevents her from winning by lines (iii)-(vi) of (4).

Therefore, by contradiction to the assumption that JULIET has a winning strategy, $v$ may contain no increment errors.

It remains to be shown that the tiling encoded by $v$ contains no vertical errors in column $n_f$.

Again, suppose for contradiction's sake that there is a vertical error in column $i = n_f$, i.e. that $v$ contains a substring of the form $uc(i)(U\{0,1\}^n)^*\#(U\{0,1\}^n)*u'c(i)$ with $(u, u') \notin V$. In this case, ROMEO has a winning strategy. As argued above, JULIET has to call every symbol from $U$ in $v$, to which ROMEO keeps returning symbols from $\hat{U}$, except for $u$ and $u'$, where ROMEO returns $u^V$ (respectively $u'^V$).

Since according to this strategy, ROMEO will never return @ or @$_N$ to a call as long as JULIET keeps calling in sequence (and JULIET loses as described above if she doesn't), JULIET cannot win by parts (1) or (3)-(5) of $T$. Since ROMEO only returns exactly two symbols from $U^V$ in subsequent lines (and therefore the rewritten string can never be in $\hat{C}$), parts (2), (6), (7) or (9) can not be reached, either. The fact that the two marked tiles $\hat{u}, \hat{u}'$ indeed violate a vertical constraint prevents JULIET from winning by (8). All that is left to show is that JULIET cannot win by part (10) of $T$.

Winning by (10) requires JULIET to call exactly one of the bits of an occurrence of $c(i)$ right after either $u^V$ or $u'^V$. Let us assume without loss of generality that this is the $k$-th bit of $c(i)$ and that its value is 0. Now, ROMEO wins by replacing it with either $\hat{0}$ or $0_N$. This is because the $[((\Sigma_P \setminus \{\#\})^n + (\Sigma_P^*\#\Sigma_P^*)_{=n+1})(0+1)]^*1$ part in (10) requires that the $k$-th bit of some $n$-bit substring in the input string does not match the $k$-th bit of $c(i)$, and since there may be at most $n-1$ bits between this bit and the final $h$ or $h'$, the string thus compared to $c(i)$ has to be the final $c_f = c(i)$. As no differing bit can be found, JULIET has no way of winning by (10), either, and therefore, ROMEO has a winning strategy. This yields the desired contradiction, so $v$ encodes a tiling without any vertical constraint violation in column $c_f$. □

We now go on to show that (a) there is a valid tiling for instance $\mathcal{I}$ if and only if (b) safe$_{L2R^+}(G) \setminus$ safe$_{L2R}(G) \neq \emptyset$.

"(a) ⇒ (b)":

Assume there is a valid tiling for $\mathcal{I}$ and let $v$ be the string encoding one such tiling. We argue that JULIET can win with a L2R$^+$ but not with a L2R strategy on $gvf$.

For a winning L2R$^+$ strategy, JULIET first calls the final $f$. If ROMEO responds with @, JULIET wins by part (1) of $T$ since $v$ encodes a valid tiling and therefore $v \in C$. Otherwise, $f$ is replaced by an $n$-bit binary number $c_f$ followed by either $h$ or $h'$. If ROMEO chose to end with $h'$, JULIET next calls the initial $g$, otherwise she plays Read on it. In the former case, JULIET plays a L2R pass on $gvc_fh$, in the latter case on $g'vc_fh'$, and by Lemma 19, JULIET has a winning L2R strategy in both of these cases because $v$ contains no vertical errors.

To show that JULIET does not have a winning L2R strategy on $gvf$, observe first that no word in $T$ ends with $f$, and therefore JULIET has to call the final $f$ at some point. If doing so is her first (and therefore only) move, she loses if ROMEO returns a string ending with $h$ or $h'$, since $T$ contains no string $gvh$ or $gvh'$ where $v$ contains exclusively symbols from $\Sigma_B$.

For similar reasons, so long as ROMEO never protests using @ or @$_N$, JULIET is forced to call all symbols from $U$ in $v$. ROMEO's winning strategy, here, is to simply return symbols from $\hat{U}$ on every call on a symbol from $U$ and answer calls on bits by returning a corresponding bit from $\hat{B}_N$. This eventually transforms $v$ into a string $v' \in \hat{C}$ (or causes JULIET to lose the game, if she calls anything other than symbols from $U$).

Finally, JULIET has to call $f$. To this, ROMEO replies with an arbitrary $n$-bit string $c_f$ and $h'$ if JULIET hasn't called the initial $g$, or $h$ if she has replaced it by $g'$. Since the only strings whose middle part $v'$ contains only symbols from $\Sigma_P$ are those from part (2) of $T$, JULIET then loses the game on $gv'c_fh'$ or $g'v'c_fh$. As we have shown that ROMEO can always win against a L2R strategy on $gvf$, it holds that $gvf \notin$ safe$_{L2R}(G)$.

"(b) ⇒ (a)":

Let $w \in$ safe$_{L2R^+}(G) \setminus$ safe$_{L2R}(G)$. We begin with some observations on the structure of $w$.

From the construction of $T$ and the rules in $R$, it follows that $w$ must begin with $g$ or $g'$ and end with $f, h, h'$ or @. If $w$ ends with @, then $w = gv@$ with $v \in C$, because no string $v$ containing symbols not in $\Sigma_B$ can be rewritten into $C$; however, in this case JULIET already has a trivial winning L2R strategy on $w$. Therefore, $w$ can not end with @ and is of the form $w = xvy$ with $x \in \{g, g'\}, y \in \{f, h, h'\}$. Also, $v$ may not contain any of the symbols $\{f, g, g', h, h'\}$, or $w$ cannot be rewritten into $T$ at all.

If $w$ ends with $h$ or $h'$, if JULIET has a winning L2R$^+$strategy on $w$, then she also has a winning L2R strategy. To prove this, assume that JULIET has a winning L2R$^+$strategy $\sigma$ on $w$, and let $s$ be the symbol on which JULIET plays her initial Call. By the replacement rules of $G$, $s \in U \cup \{g, 0, 1\}$.

- $s = g$: In this case, JULIET's first Call takes place on the first letter of $w$, since (as stated above) $g$ may appear nowhere else in the input string. Therefore, $\sigma$ is already a L2R strategy, since no left step is necessary to start the L2R pass after a single Call on the initial symbol of $w$.

- $s \in U \cup \{0, 1\}$: In this case, JULIET may not call any symbol in $v$ to the left of $s$ after calling $s$, because ROMEO can always return @ on this second call and $T$ contains no strings with an @ to the left of a symbol in $\hat{u} \cup U^V \cup \{\hat{0}, \hat{1}, 0_N, 1_N, @, @_N\}$. Therefore, the only symbol in $w$ left of $s$ which JULIET may call after calling $s$ is an initial $g$. Doing so can only be part of a winning strategy if $w$ ends with $h'$; however, in that case, $\sigma$ can be transformed into an equivalent L2R strategy by calling $g$ before $s$.

Since $w \in$ safe$_{L2R^+} \setminus$ safe$_{L2R}$, this implies that $w$ cannot end with $h$ or $h'$.

If $w$ ends with $f$, then $w = gvf$ with $v \in C$. This is the case because no word in $T$ ends with $f$, so JULIET inevitably has to call the final $f$ and ROMEO can always win on words not of the structure $gvf$ by replacing $f$ with @.

By the same argument, JULIET's initial call has to be on the final $f$; if she doesn't start by calling $f$, ROMEO can later on reply to the necessary call on $f$ with @. After the initial call on $f$, play proceeds with a left-to-right pass over a string of the form $gvc_fh$ or $gvc_fh'$.

On a string of the form $gvc_fh'$, JULIET's first move in her L2R pass has to be a Call on $g$, since (as argued above) ROMEO may always replace tiles from $U$ in $v$ by marked tiles from $\hat{U}$ and the only strings in $T$ ending with $h'$ and containing only symbols from $\Sigma_P$ in $v$ are of the form $g'vc_fh'$.

Therefore, JULIET has a winning strategy on a string of the form $gvc_fh$ or $g'vc_fh'$, and by Lemma 19, this is the case if and only if $v$ encodes a tiling candidate with no vertical errors in the column $n_f$ encoded by $c_f$ and no increment errors. Since ROMEO is free to chose any column index string $c_f \in \{0,1\}^n$ and by our prerequisite JULIET has a winning strategy for any of these, $v$ may not contain vertical errors in any column and therefore encodes a valid tiling.

This concludes our proof that the existence of a word on which JULIET has a L2R$^+$ but no L2R winning strategy in $G$ implies the existence of a valid tiling for $\mathcal{I}$. □

LEMMA 20. *For a number $i \in \{0,1,\ldots,2^n - 1\}$ let $c(i)$ be the n-bit binary encoding of $i$, and for an n-bit string $c$ let $c_k$ denote the k-th position of $c$. For any two numbers $i, j \in \{0, 1, \ldots, 2^n - 1\}$, it holds that $j \neq i + 1$ if and only if there exists a number $k \leq n$ such that one of the following conditions holds:*

(a) $c(i)_k \neq c(j)_k$, $c(i) \neq 1^n$ and for some $k' > k$, it holds that $c(i)_{k'} = 0$ or $c(j)_{k'} = 1$;

(b) $c(i)_k = c(j)_k$ and it holds that either $k = n$ or $c(i)_{k+1} = 1$ and $c(j)_{k+1} = 0$

(c) $c(i) = 1^n$

PROOF. ($\Rightarrow$) Let $c(i), c(j)$ be as described with $j \neq i+1$. If $i = j$, then the first part of (b) holds with $k = n$.

If $i + 1 < j$, then let $k$ be 1 plus the length of the longest common prefix of $c(i)$ and $c(j)$ (i.e. $k$ is the smallest index such that $c(i)_k \neq c(j)_k$). Since $i + 1 < j \leq 2^n - 1$, $c(i) \neq 1^n$, $c(i)_k = 0$ and $c(j)_k = 1$. If $c(i)_{k'} = 1$ and $c(j)_{k'} = 0$ were to hold for all $k' > k$ then it would follow that $i + 1 = j$, so condition (a) must be fullfilled.

If $i > j$ and $c(i) = 1^n$, then (c) holds. Let therefore $c(i) \neq 1^n$. If $c(i)_1 \neq c(j)_1$, then $c(i)_1 = 1$ and $c(j)_1 = 0$; however, since $c(i) \neq 1^n$, there exists a $k'$ with $c(i)_{k'} = 0$, and therefore (a) holds with $k = 1$. Otherwise, let $k$ be the length of the longest common prefix of $c(i)$ and $c(j)$. Then, since $i > j$, it holds that $c(i)_k = c(j)_k$ and $c(i)_{k+1} = 1, c(j)_{k+1} = 0$, and thus (b) holds.

($\Leftarrow$) Assume that $i + 1 = j$ for a given substring of $v$ of the form $c(i)Uc(j)$. Then $c(i) \neq 1^n$, since $i < j \leq 2^n - 1$, and therefore (c) cannot hold. Let $k \leq n$.

If $c(i)_k = 0$ and $c(j)_k = 1$, then $k = n$; if $c(i)_k = 1$ and $c(j)_k = 0$, then $k < n$. In both cases, since $i + 1 = j$, it follows that $c(i)_{k'} = 1$ and $c(j)_{k'} = 0$ for all $k' > k$, and therefore (a) cannot hold.

If $c(i)_k = c(j)_k$, then either $c(i)_{k+1} = c(j)_{k+1}$ or (again because $i + 1 = j$) $c(i)_{k+1} = 0, c(j)_{k+1} = 1$. In any of these cases, (b) does not hold. □

PROPOSITION 21. *L2RALL is hard for EXPTIME, even for games with finite replacement language.*

PROOF. The proof is by a polynomial time reduction from the L2R word problem, i.e., given a game $G = (\Sigma, R, T)$ and a string $u$, decide whether $u \in \text{safe}_{L2R}(G)$. This problem is was shown to be EXPTIME-complete in [8].

To this end, we show how to construct in polynomial time a game $G' = (\Sigma', R', T')$ from $G$ and $u$ such that the following statements are equivalent.

(a) $u \in \text{safe}_{L2R}(G)$.

(b) $\text{safe}(G') \setminus \text{safe}_{L2R}(G') \neq \emptyset$.

The construction of $G'$ will ensure that JULIET can deduce a winning strategy on a string $g_0 u h_0$ wrt $G'$ with a single Call move in a first phase followed by an L2R phase if and only if she has an L2R winning strategy on $u$ in $G$. In $G'$ we use additional symbols $g_0, g_1, g_2, h_0, h_1, h_2, \#, @$, where

- $g_0, g_1, g_2, h_0, h_1, h_2$ are used to rule out L2R strategies for many strings,

- @ can be used by ROMEO to "protest" if JULIET deviates from the intended flow of the game, and

- # is used to force JULIET to follow an L2R strategy on $u$ (or otherwise ROMEO can "protest").

The alphabet $\Sigma'$ is $\Sigma \cup \{g_0, g_1, g_2, h_0, h_1, h_2, \#, @\}$ and we assume that the latter eight symbols do not belong to $\Sigma$.

For each rule $f \to w_1 \mid \cdots \mid w_\ell$ of $R$, there is a rule $f \to \#w_1 \mid \cdots \mid \#w_\ell \mid @$ in $R'$. Furthermore, $R'$ contains the following rules.

- $g_0 \to g_1 \mid @$

- $g_1 \to g_2 \mid @$

- $h_0 \to h_1 \mid h_2 \mid @$

For a string $w \in (\Sigma \cup \{\#\})^*$, we write $\text{cl}(w)$ for the string that results from $w$ by eliminating all occurrences of $\#$.

The target language $T'$ of $G'$ contains

- all strings $g_1 w h_1$ with $\text{cl}(w) \in T$;

- all strings $g_2 w h_2$ with $\text{cl}(w) \in T$;

- the string $g_0 u @$;

- all strings of the form $gwh$ where $g \in \{g_0, g_1, g_2\}$, $h \in \{h_0, h_1, h_2\}$, and in $w$ there is at least one occurrence of @ but no occurrence of $\#$ to the right of an occurrence of @;

- all strings $@wh_1$ and $@wh_2$, where $w$ only contains symbols from $\Sigma$.

Clearly, $G'$ can be constructed in polynomial time from $G$ and $u$, in particular a DFA for $T'$ (assuming a DFA for $T$).

It remains to show that (a) and (b) are indeed equivalent.

"(a) $\Rightarrow$ (b)":

We show that if $u \in \text{safe}_{L2R}(G)$ it follows that $g_0 u h_0 \in \text{safe}(G') \setminus \text{safe}_{L2R}(G')$.

First $g_0 u h_0 \in \text{safe}(G')$ as JULIET can choose the last position (carrying $h_0$) first. If ROMEO answers with @ she immediately wins as $g_0 u @ \in T'$. Otherwise, she enforces $g_1$ as first symbol if ROMEO chose $h_1$ and $g_2$ if ROMEO chose $h_2$. If ROMEO chooses @ for the first symbol, JULIET wins directly as strings of the forms $@w h_1$ and $@w h_2$ with $w \in \Sigma^*$ are in $T'$. Then JULIET can basically follow her L2R winning strategy on $u$. It is easy to see that she wins the game in this fashion.

We show next that $g_0 u h_0 \notin \text{safe}_{L2R}(G')$. Clearly, JULIET needs to play a Call on the last position as she cannot enforce a win otherwise (ROMEO simply never protests). However, ROMEO can reply by $h_1$ just if the first position of the string is not $g_1$, enforcing a win for ROMEO.

Thus, (a) $\Rightarrow$ (b).

"(b) $\Rightarrow$ (a)":

Assume that there is a word $v \in \text{safe}(G') \setminus \text{safe}_{L2R}(G')$. We start with some observations on what $v$ can look like.

1. The word $v$ must begin with some $g_i \in \{g_0, g_1, g_2\}$. Indeed, no letter not in $\{g_0, g_1, g_2\}$ can ever be rewritten into a letter in $\{g_0, g_1, g_2\}$ and the only strings that are accepted that do not begin with such a letter are strings on the form @$wh_1$ or @$wh_2$. If JULIET wins on a string that begins with @, it must be by never playing Call, since otherwise ROMEO could protest with a second @ and win. Thus, if JULIET wins, she wins with an L2R-strategy.

2. The word $v$ must end with some $h_j \in \{h_0, h_1, h_2\}$. There is only a single accepted string that is accepted that does not end with such a letter and no other letter can be rewritten into them. If the string $v$ ends with @, it is either $g_0u$@, in which case JULIET wins with an L2R-strategy by just reading it, or it another string, in which case she cannot win, since any Call move can be answered by ROMEO with a second @ symbol.

3. If JULIET has a winning strategy on $g_iwh_j$, then the strategy must play left to right on $g_iw$. If JULIET plays Call on a symbol $f$ in $w$, ROMEO can answer with a string $\#w_j \in R_f$, introducing a $\#$ symbol into the word. If JULIET ever plays a Call on a position to the left of this symbol, ROMEO can protest with an @ symbol, creating a word with an @ to the left of a $\#$. After this, JULIET cannot win.

4. That JULIET can win on $v = g_iwh_j$, but not with an L2R-strategy means that she needs to call on the last position before completing play on $g_iw$. This implies $h_j = h_0$.

5. On strings of the form $g_1wh_0$ or $g_2wh_0$, JULIET has no winning strategy. Indeed, since no accepted string ends with $h_0$, JULIET will sooner or later have to play Call on the last position. When she does this, ROMEO can answer with @. The only string ending with @ that is accepted is $g_0u$@, but $g_1$ or $g_2$ can never be rewritten into $g_0$, so JULIET loses.

From (1)–(5) above, we can conclude that if $v \in \text{safe}(G') \setminus \text{safe}_{L2R}(G')$, then $v$ has the form $g_0wh_0$. When JULIET starts play on a word $g_0wh_0$ by calling on the last position, ROMEO can answer with @. The only accepted string that ends with @ is $g_0u$@. This means that the string $v$ must be $g_0uh_0$. ROMEO can, however, also answer the call on $h_0$ with $h_1$ or $h_2$. In this case, JULIET must play an L2R-strategy that transforms $g_0u$ into some $g_iw'$ with $\text{cl}(w') \in T$. This same strategy, restricted to $u$ and ignoring the $\#$-symbols, is a winning L2R-strategy on $u$ in $G$. □

## 8. CONCLUSION

We investigated a practically relevant restriction of strategies for context-free games and their relation to general strategies. That L2RALL is EXPSPACE-complete in general but EXPTIME-complete in the restricted case where the replacement languages in $G$ are finite, is somewhat surprising, since the word problem of checking whether a given string is safely rewritable in a left-to-right fashion is EXPTIME-complete in both cases[8].

The automaton construction for $\text{safe}_{L2R}$ we give here can be generalised to yield automata for strings which can be safely rewritten using up to $k$ left steps (with a full L2R pass being played before each left step). This is done by generalising our definition of effects to $k$-effects, each of which is a set of sets of $(k-1)$-effects representing games on later passes. In this framework, effects as defined in this paper would correspond to 1-effects.

It can also be shown that for every game $G$ there is a game $G'$ with finite replacement languages whose safely rewritable strings are exactly those of $G$.

A further open frontier remains in the form of *One-Pass* (1P) strategies [2], which restrict L2R strategies by forcing JULIET to make her decisions in a streaming manner, i.e. without knowing the entire input string. While Abiteboul, Milo and Benjelloun [2] have shown a number of interesting properties of such strategies, the general problem of testing whether every safely L2R-rewritable string of a given game can also safely rewritten in a 1P fashion is not even known to be decidable.

## 9. REFERENCES

[1] Serge Abiteboul, Angela Bonifati, Gregory Cobena, Ioana Manolescu, and Tova Milo. Dynamic XML documents with distribution and replication. In Alon Y. Halevy, Zachary G. Ives, and AnHai Doan, editors, *SIGMOD Conference*, pages 527–538. ACM, 2003.

[2] Serge Abiteboul, Tova Milo, and Omar Benjelloun. Regular rewriting of active XML and unambiguity. In *PODS*, pages 295–303, 2005.

[3] ActiveXML. http://www-rocq.inria.fr/verso/Gemo/Projects/axml.

[4] B.S. Chlebus. Domino-tiling games. *Journal of Computer and System Sci.*, 32(3):374–392, 1986.

[5] E. Grädel, W. Thomas, and T. Wilke, editors. *Automata, Logics, and Infinite Games. A Guide to Current Research.* Springer, 2002.

[6] J. Kulbatzki. Active XML und Untersuchungen perfekter Rewriting-Strategien für kontexfreie Spiele auf Strings. Master's thesis, Technische Universität Dortmund, 2010.

[7] Tova Milo, Serge Abiteboul, Bernd Amann, Omar Benjelloun, and Frederic Dang Ngoc. Exchanging intensional XML data. *ACM Trans. Database Syst.*, 30(1):1–40, 2005.

[8] Anca Muscholl, Thomas Schwentick, and Luc Segoufin. Active context-free games. *Theory Comput. Syst.*, 39(1):237–276, 2006.

[9] Walter J. Savitch. Relationships between nondeterministic and deterministic tape complexities. *J. Comput. Syst. Sci.*, 4(2):177–192, 1970.

[10] P. van Emde Boas. The convenience of tilings. In *Complexity, Logic and Recursion*, volume 187 of *Lec. Notes in Pure and App. Math.*, pages 331–363. Routledge, 1997.